\documentclass[aps,prl,twocolumn,superscriptaddress]{revtex4-1}
\usepackage{amsmath,amssymb}
\usepackage{graphicx}
\usepackage{subfigure}
\usepackage{verbatim}
\usepackage{amsfonts}
\usepackage{dcolumn}
\usepackage{bm}
\usepackage{color}
\usepackage[colorlinks,citecolor=blue]{hyperref}

\usepackage{mathrsfs}

\newcommand{\beq}{\begin{eqnarray} }
\newcommand{\eeq}{\end{eqnarray} }
\newcommand{\Beq}{\begin{eqnarray*} }
\newcommand{\Eeq}{\end{eqnarray*} }
\newcommand{\Bmat}{\left(\begin{matrix}}
\newcommand{\Emat}{\end{matrix}\right)}

\begin{document}

\title{One Proximate Kitaev Spin Liquid in the $K$-$J$-$\Gamma$ Model
on the Honeycomb Lattice}

\author{Jiucai Wang}
\affiliation{Department of Physics, Renmin University of China, Beijing 100872, China}

\author{B. Normand}
\affiliation{Neutrons and Muons Research Division, Paul Scherrer Institute,
CH-5232 Villigen-PSI, Switzerland}

\author{Zheng-Xin Liu}\email{liuzxphys@ruc.edu.cn}
\affiliation{Department of Physics, Renmin University of China, Beijing 100872, China}

\date{\today}

\begin{abstract}
In addition to the Kitaev ($K$) interaction, candidate Kitaev materials
also possess Heisenberg ($J$) and off-diagonal symmetric ($\Gamma$) couplings.
We investigate the quantum ($S = 1/2$) $K$-$J$-$\Gamma$ model on the honeycomb
lattice by a variational Monte Carlo (VMC) method. In addition to the
``generic'' Kitaev spin liquid (KSL), we find that there is just one
proximate KSL (PKSL) phase, while the rest of the phase diagram contains
different magnetically ordered states. The PKSL is a gapless Z$_2$ state with
14 Majorana cones, which in contrast to the KSL has a gapless spin response.
In a magnetic field applied normal to the honeycomb plane, it realizes two of
Kitaev's gapped chiral spin-liquid phases, of which one is non-Abelian with
Chern number $\nu = 5$ and the other is Abelian with $\nu = 4$. These two
phases could be distinguished by their thermal Hall conductance.
\end{abstract}

\maketitle

The Kitaev model \cite{c0} of bond-dependent Ising-type spin interactions
on the honeycomb lattice offers exactly soluble examples of both gapped and
gapless quantum spin liquids (QSLs). The magnetically disordered ground states
of different QSLs are the consequence of strong intrinsic quantum fluctuations
and provide particularly clean realizations of different fundamental phenomena.
The gapped Kitaev QSL has Z$_2$ Abelian topological order, while the
quintessential ``Kitaev spin liquid'' (KSL) is the gapless state whose
low-energy excitations form two Majorana cones, whereas its Z$_2$ flux
excitations are gapped. In an applied magnetic field, the Majorana cones
become gapped and the resulting state is a chiral spin liquid (CSL) with
Ising-type non-Abelian anyonic excitations, which have potential application
in fault-tolerant topological quantum computation.

Thus the experimental realization of the Kitaev model has moved to the
forefront of research in strongly correlated materials. While transition-metal
compounds with strong spin-orbit coupling do realize Kitaev-type interactions
\cite{rjk,rcjk}, these ``candidate Kitaev'' materials typically possess
in addition significant non-Kitaev interactions, which lead to Na$_2$IrO$_3$
\cite{ryea,rchoietal} and $\alpha$-RuCl$_3$ \cite{rfgft,rsea,rjea,rcaoetal}
exhibiting magnetic order at low temperatures. Although H$_3$LiIr$_2$O$_6$
\cite{rhliiro} is not ordered, it appears to show strong impurity and
extrinsic disordering effects. At the same order in a strong-coupling
treatment \cite{c1}, the Kitaev ($K$) interaction is accompanied by
Heisenberg ($J$) and off-diagonal symmetric ($\Gamma$) interactions, and
thus the focus of the field has become the understanding of ``proximate
Kitaev'' physics in this class of model, also under applied magnetic fields
\cite{rlea,rbea,rzea,rins_field} and pressures \cite{rucl3_pressure}.

In this Letter, we investigate the $K$-$J$-$\Gamma$ extended Kitaev model by
variational Monte Carlo (VMC) studies of a spinon representation. Guided by
the projective symmetry group (PSG), we obtain the global $K$-$J$-$\Gamma$
phase diagram and show that it contains two distinct QSL phases among several
classically ordered phases. One QSL, at small $J$ and $\Gamma$, is the generic
KSL. At larger $\Gamma$ we find one proximate KSL (PKSL), a non-Kitaev QSL
sharing the same PSG as the KSL but with 14 Majorana cones in the first
Brillouin zone and gapless spin excitations. In an applied ${\hat c}$-axis
field, all 14 cones are gapped and the PKSL hosts two exotic CSL phases, one
with non-Abelian anyons and one with Abelian anyons. These results shed
crucial new light on the parameter-space constraints and (induced) spin-liquid
phases of candidate Kitaev materials.

In the candidate materials known to date, $K < 0$ is believed to be
ferromagnetic \cite{rwljv,rins,c4,rwdyl}, while $J$ has been argued to have
both signs and extended nature, but all studies of the $\Gamma$ (and $\Gamma'$)
term(s) take $\Gamma > 0$ \cite{rjav2,rcm}. The role of $\Gamma$ in a fully
quantum model remains little studied \cite{c2}, but from classical models
$\Gamma$ is thought to explain the strongly anisotropic field response of
$\alpha$-RuCl$_3$ \cite{rjea,rjav2}. In general, it is not yet accepted that
a $J = 0$ (i.e.~$K$-$\Gamma$) model can support a magnetically ordered ground
state \cite{rjav2}, and it has been claimed on the basis of exact
diagonalization of small systems \cite{c1,rcywkk} and iDMRG studies of
narrow cylinders \cite{c3} that multiple QSL phases may exist in the
$K$-$J$-$\Gamma$ model at small $J$.

The model we consider is
\begin{equation}\label{KJG}
H = \sum_{\langle i,j \rangle \in \gamma} K S_i^\gamma S_j^\gamma + J {\vec S}_i
\cdot {\vec S}_j + \Gamma (S_i^\alpha S_j^\beta + S_i^\beta S_j^\alpha),
\end{equation}
where $\langle i,j \rangle$ denotes nearest-neighbor sites and $\gamma$
both the bond type on the honeycomb lattice and the spin index. Due to the
spin-orbit coupling, the symmetry group of the model, $G = D_{3d} \times Z_2^T$
where $Z_2^T = \{E,T\}$ is time reversal, is finite, i.e.~its elements are
discrete operations combining the space-time and spin degrees of freedom.

We find the ground state of the $S = 1/2$ $K$-$J$-$\Gamma$ model by VMC,
which is a powerful method for the study of spin-liquid phases.
We first introduce the fermionic slave-particle representation $S_i^m =
\frac{1}{2} C_i^\dagger \sigma^m C_i$, where $C_i^\dagger = (c_{i\uparrow}^\dagger,
c_{i\downarrow}^\dagger)$, $m \equiv x,y,z$, and $\sigma^m$ are Pauli matrices.
The particle-number constraint, $\hat{N_i} = c_{i\uparrow}^\dagger c_{i\uparrow} +
c_{i\downarrow}^\dagger c_{i\downarrow} = 1$, should be imposed at every site to
ensure that the size of the fermion Hilbert space is the same as that of
the physical spin. This complex fermion representation is equivalent to
the Majorana representation introduced by Kitaev \cite{c0}. It has a local
SU(2) symmetry that is independent of the SU(2) spin-rotation operations
and can be considered as a gauge structure \cite{AndersonSU2}, as discussed
in Sec.~S1A of the Supplemental Material (SM) \cite{sm}. The spin interactions
in Eq.~(\ref{KJG}) are rewritten in terms of interacting fermionic operators
and decoupled into a noninteracting mean-field Hamiltonian (Sec.~S1B of the
SM \cite{sm}), which for the most general spin-orbit-coupled spin liquid
has the form
\begin{eqnarray}
H_{\rm mf} = & \!\sum_{\langle i,j \rangle\in\gamma} {\rm Tr} \, [U_{ji}^{(0)} \psi_i ^\dag
\psi_j] + {\rm Tr} \, [U_{ji}^{(1)} \psi_i^\dag (iR_{\alpha\beta}^\gamma) \psi_j]
\nonumber \\ & \;\; + {\rm Tr} \, [U_{ji}^{(2)} \psi_i^\dag \sigma^\gamma \psi_j]
 + {\rm Tr} \, [U_{ji}^{(3)} \psi_i^\dag \sigma^\gamma R_{\alpha\beta}^\gamma \psi_j],
\label{emfh}
\end{eqnarray}
where $\psi_i = (C_i\ \bar C_i)$, $\bar C_i = (c_{i\downarrow}^\dag, -
c_{i\uparrow}^\dag)^T$, $R_{\alpha\beta}^\gamma = - \frac{i}{\sqrt{2}} (\sigma^\alpha +
\sigma^\beta)$ is a rotation matrix, and the quantities $U_{ji}^{(m)}$, with
$\gamma$ specified by $\langle i,j \rangle$, are mean-field parameters.

\begin{figure}
\includegraphics[width=8.6cm]{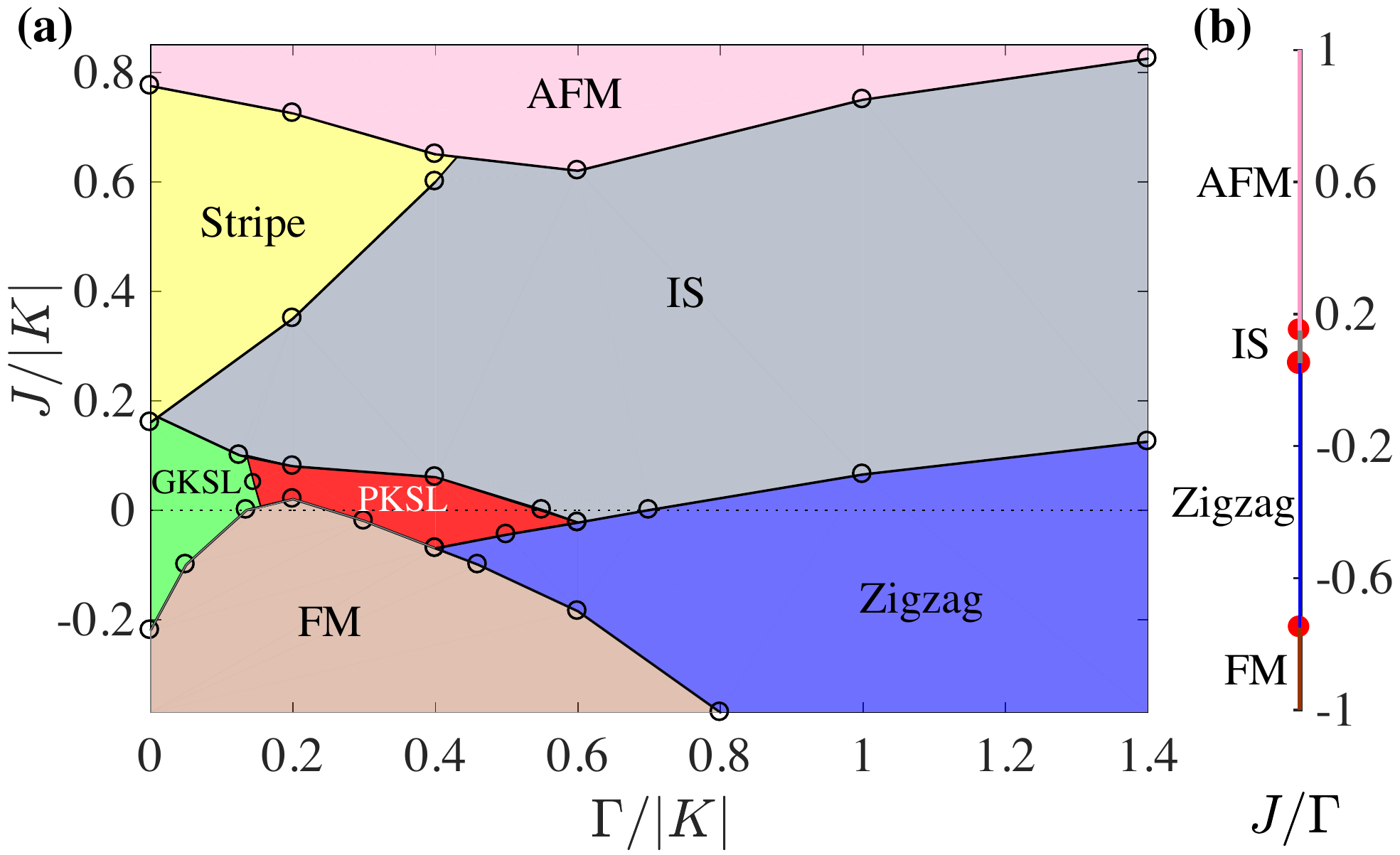}
\caption{(a) Phase diagram of the quantum $K$-$J$-$\Gamma$ model for $K < 0$
in the limit of large system size. There are two QSL phases of different types
but with the same PSG, the GKSL and the PKSL. The magnetically ordered phases are
antiferromagnetic (AFM), stripe, incommensurate spiral (IS), zigzag, and
ferromagnetic (FM) order. (b) Detail of the limit $|K|/\Gamma \to 0$, where
all phases are magnetically ordered; the transitions occur at $J/\Gamma =
0.15$, 0.05, and $- 0.75$.}
\label{Fig_phsdgrm}
\end{figure}

In the states described by Eq.~(\ref{emfh}), the SU(2) gauge symmetry is
usually reduced to U(1) or Z$_2$, which is known as the invariant gauge group
(IGG) \cite{igg}. The Majorana mean-field solution of the Kitaev model has
IGG Z$_2$. Because the KSL is a special (exactly soluble) point in the model
of Eq.~(\ref{KJG}), one expects a finite regime of QSL states connected
adiabatically to it; following Ref.~\cite{rsyb} we call this the generic
KSL (GKSL). A QSL ground state preserves the full symmetry group,
$G$, and so does the mean-field Hamiltonian. However, a general symmetry
operation of this Hamiltonian is a space-time and spin operation in $G$
combined with an SU(2) gauge transformation. These new symmetry operations
form a larger group, known as the PSG, which is equivalent to a central
extension of $G$ by the IGG (Sec.~S2 of the SM \cite{sm}). The PSG of the KSL
is known exactly \cite{c5} and the corresponding mean-field Hamiltonian must
respect it. The PSG reduces the number of independent parameters and, as
detailed in Sec.~S3 of the SM \cite{sm}, the coefficients $U_{ji}^{(m)}$ in
Eq.~(\ref{emfh}) are constrained to the forms $U_{ji}^{(0)} = i \eta_0 +
i (\rho_a + \rho_c)$, $U_{ji}^{(1)} = i (\rho_a - \rho_c  + \rho_d)
(\tau^\alpha + \tau^\beta) + i \eta_3 (\tau^x + \tau^y + \tau^z)$, $U_{ji}^{(2)}
 = i (\rho_a + \rho_c) \tau^\gamma + i \eta_5 (\tau^x + \tau^y + \tau^z)$, and
$U_{ji}^{(3)} = i (\rho_c - \rho_a - \rho_d) (\tau^\alpha - \tau^\beta)$.

In addition we allow competing magnetically ordered phases by including
the term $H_{\rm mf}' = {\textstyle \frac12} \sum_i \pmb M_i \cdot C_i^\dag
\pmb \sigma C_i$ in the mean-field Hamiltonian \cite{c2}. The ordering
pattern of $\pmb M_i$ is set from the classical solution within the
single-${\pmb Q}$ approximation, leaving only the amplitude, $M$, and
the canting angle, $\phi$, to be determined variationally (Sec.~S3 of
the SM \cite{sm}). The power of the VMC approach is that it allows the
particle-number constraint to be enforced locally, by performing Gutzwiller
projection of the mean-field ground states to obtain the trial wave functions
$|\Psi(x) \rangle = P_G |\Psi_{\rm mf}(x) \rangle$, where $x$ denotes the
variational parameters $(\rho_a,\rho_c,\rho_d,\eta_0,\eta_3,\eta_5,M,\phi)$.
These are determined by minimizing the trial ground-state energy, $E (x) =
\langle \Psi(x) |H| \Psi(x) \rangle / \langle \Psi(x)| \Psi(x) \rangle$, in
calculations performed on tori of up to 14$\times$14 unit cells, i.e.~392
lattice sites. Because the final variational states depend crucially on the
mean-field Hamiltonian, a meaningful VMC procedure requires a careful choice
and comparison of decoupling channels, and we have tested many spin-liquid
and magnetic ansatzes (Sec.~S3).

\begin{figure*}
\includegraphics[width=7.cm]{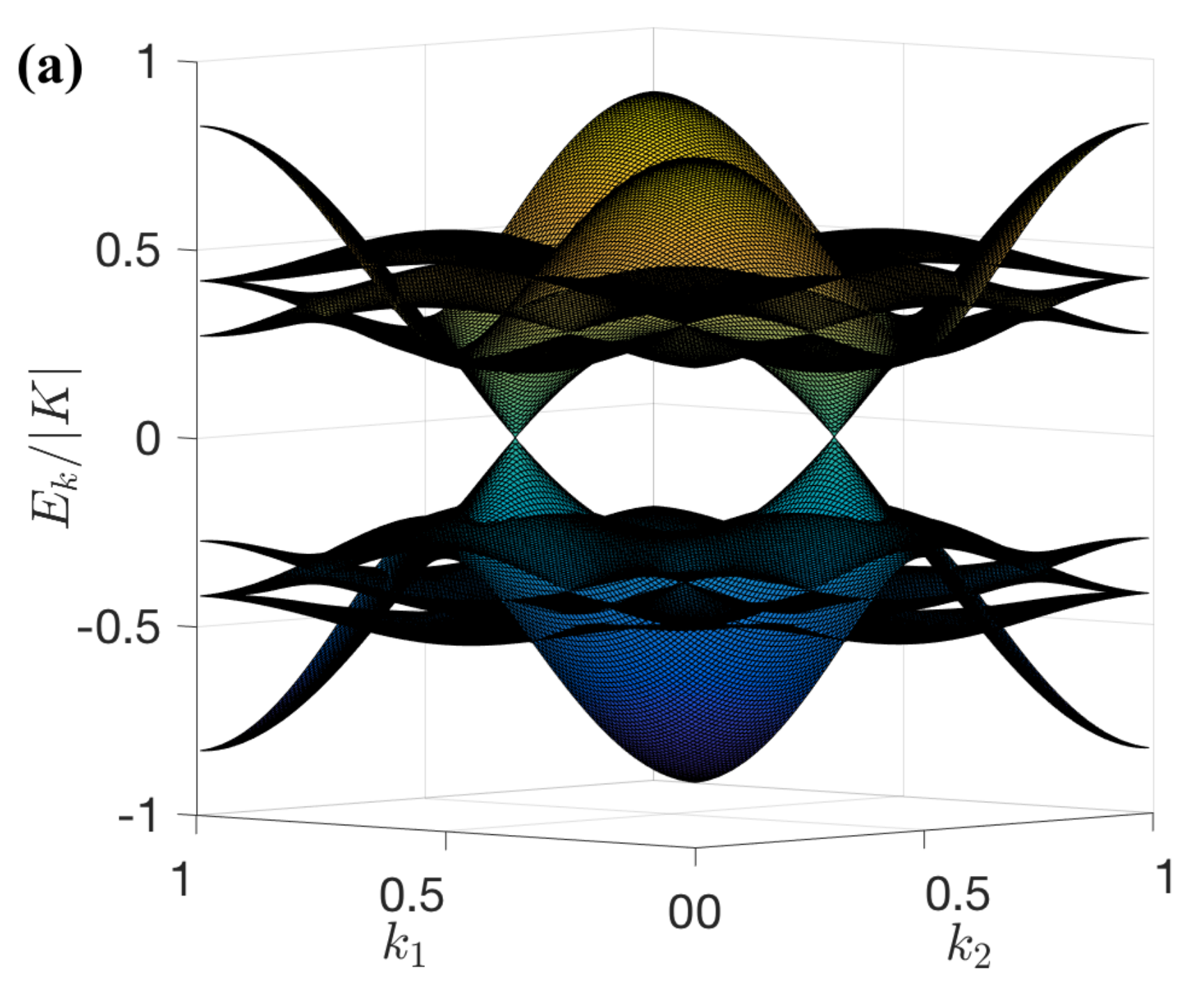}
\includegraphics[width=7.cm]{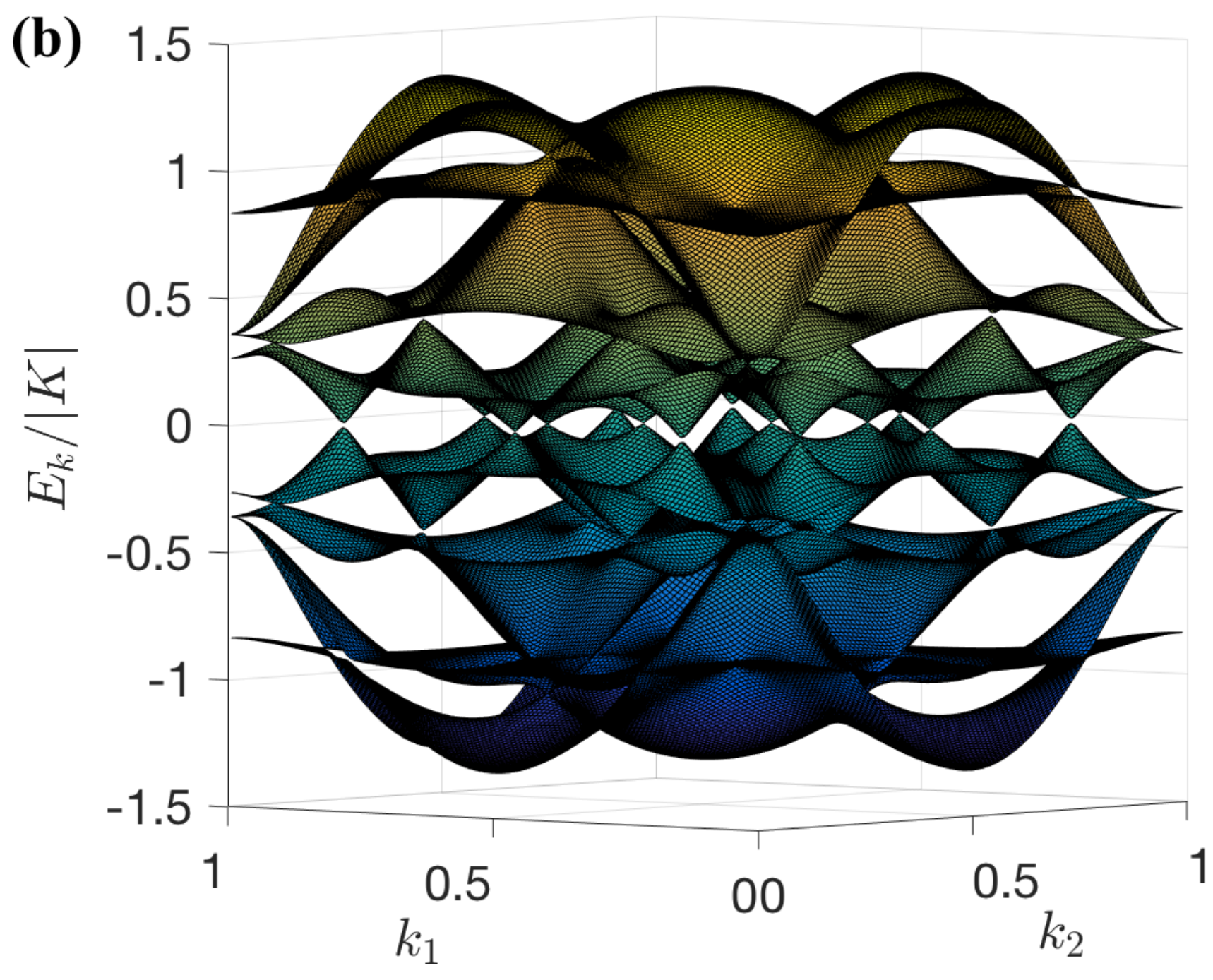}
\includegraphics[width=3.7cm]{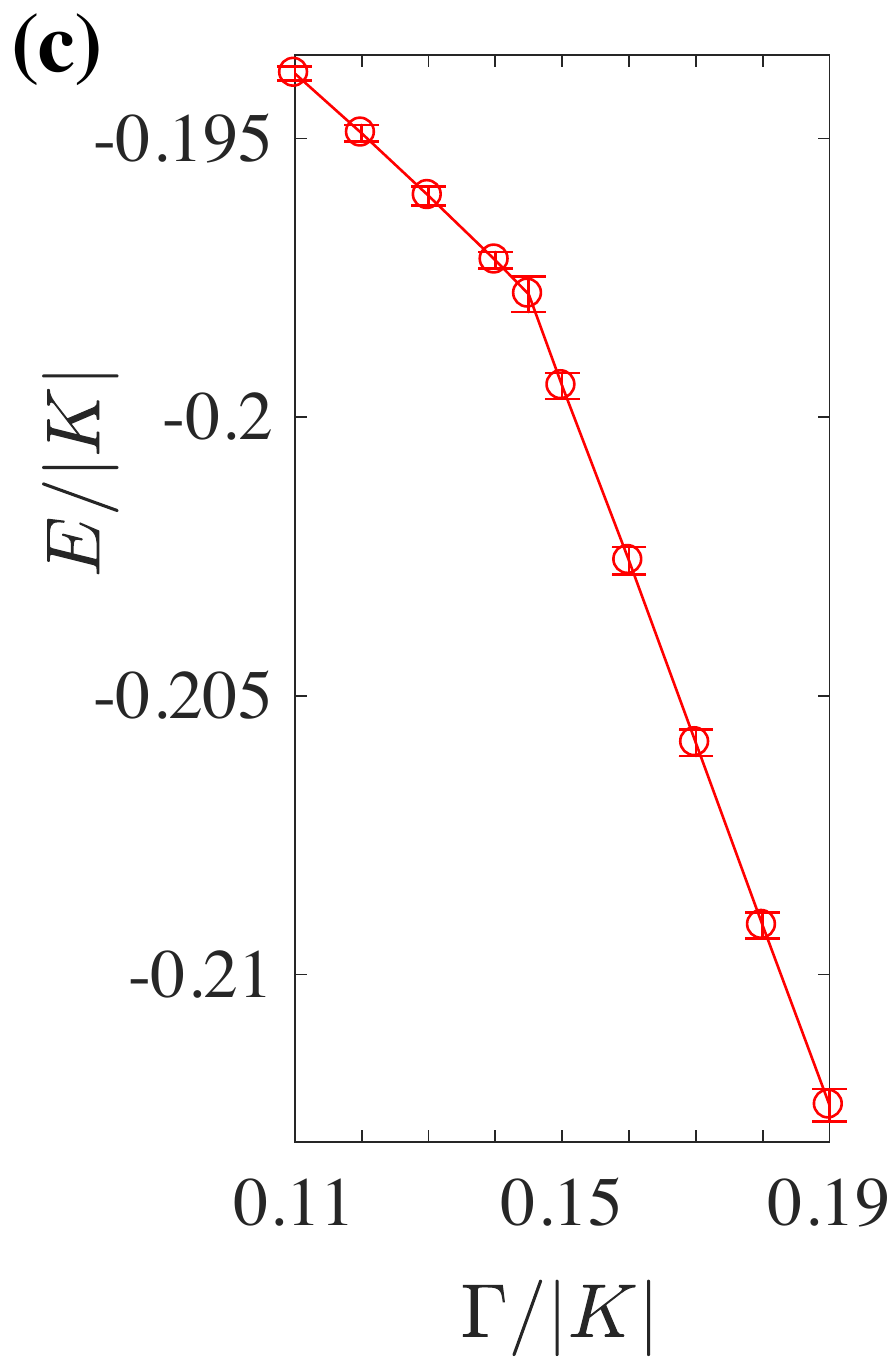}
\includegraphics[width=5.cm,height=3.4cm]{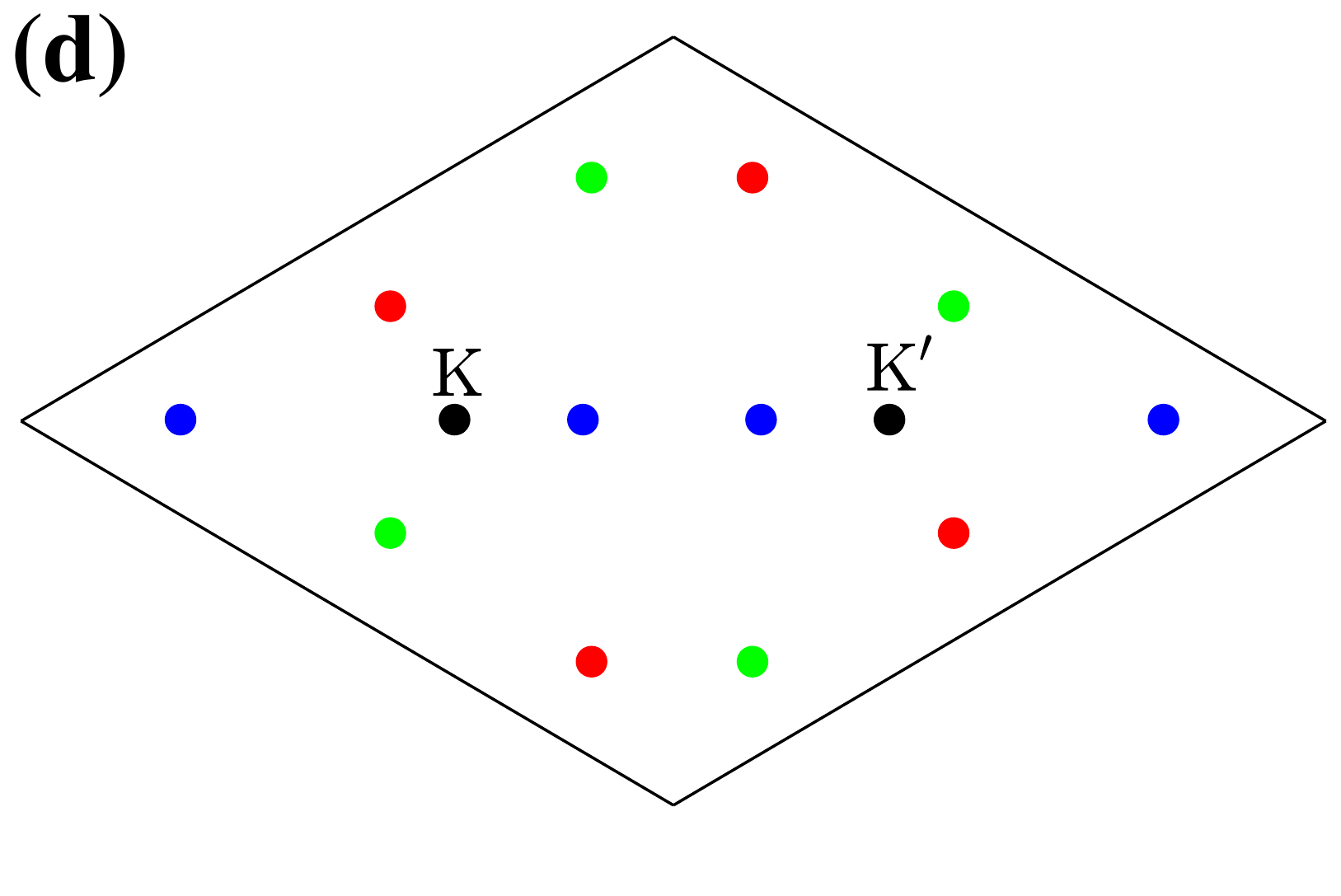}\ \ \ \ \
\includegraphics[width=5.8cm,height=3.4cm]{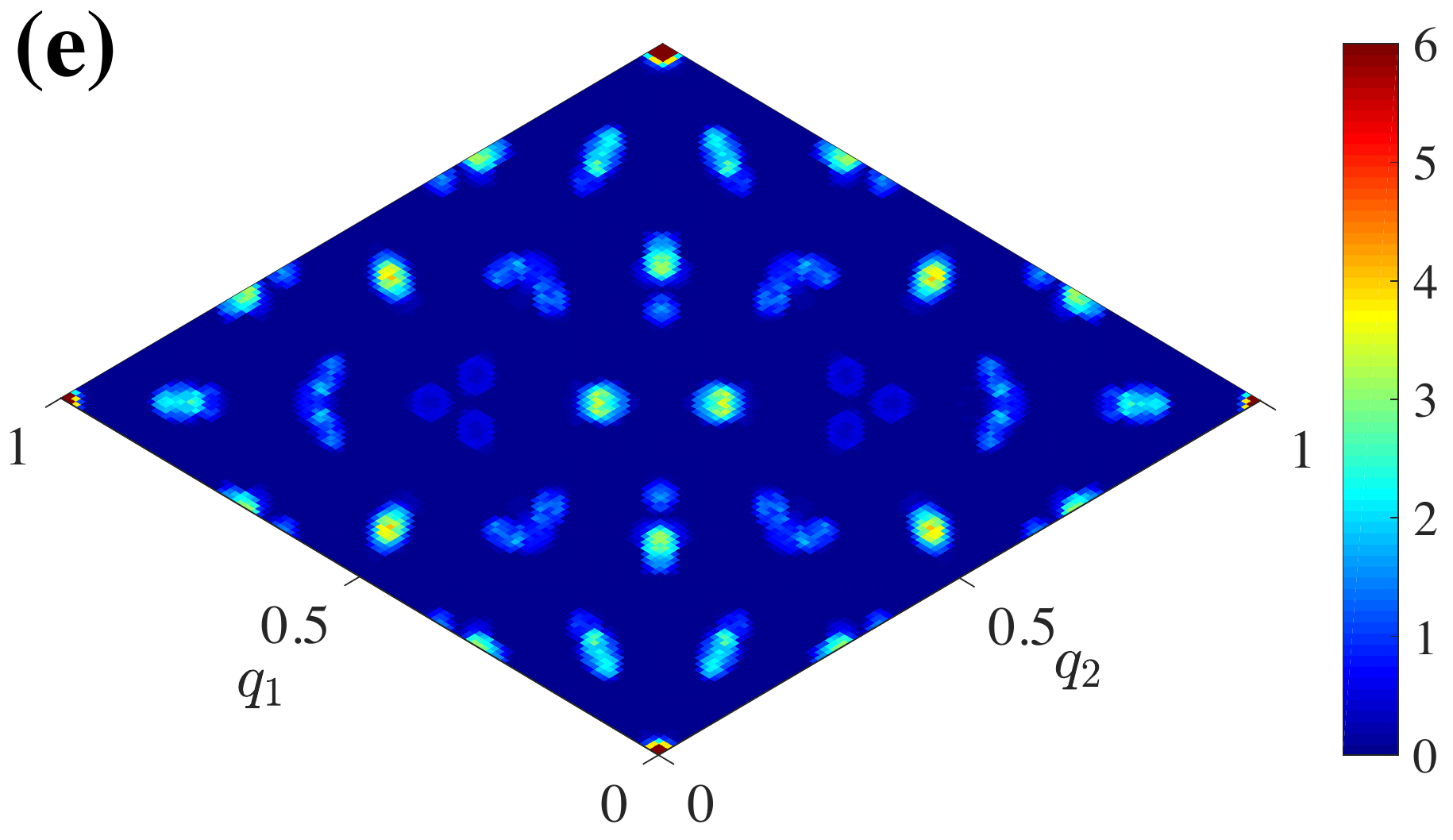}\ \ \ \ \
\includegraphics[width=5.8cm,height=3.4cm]{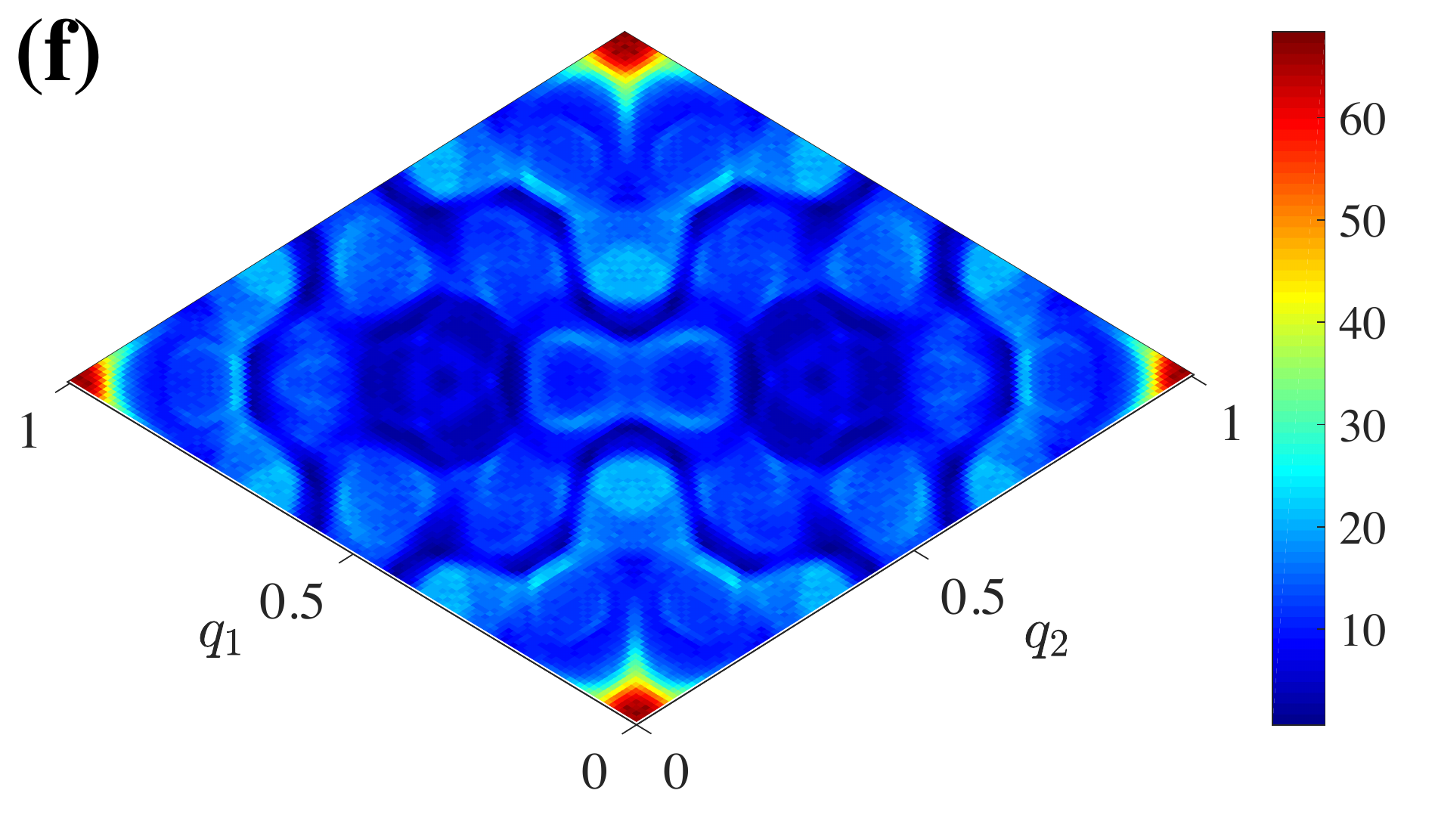}
\caption{(a) Spinon dispersion in the GKSL, drawn with $\Gamma/|K| = 0.1$
and $J = 0$, showing 2 Majorana cones. (b) Spinon dispersion in the PKSL,
drawn with $\Gamma/|K| = 0.3$ and $J = 0$, showing 14 Majorana cones.
(c) Ground-state energy per site of the $K$-$J$-$\Gamma$ model at fixed
$J/|K| = 0.05$, showing a clear first-order phase transition. (d) Locations of
the 14 cones of the PKSL in the first Brillouin zone. (e) Dynamical structure
factor of the PKSL at low energy (integrated over energies $0 \le \omega/|K|
\le 0.08$). (f) Dynamical structure factor of the PKSL at $\omega/|K| = 0.15$
(integrated over $0.13 \le \omega/|K|
\le 0.17$).}
\label{Fig_Mcones}
\end{figure*}

Figure \ref{Fig_phsdgrm} shows the VMC phase diagram of the $K$-$J$-$\Gamma$
model at zero applied field. As a benchmark, we note that our phase boundaries
at $\Gamma = 0$, $J > 0$ agree quantitatively with those of Ref.~\cite{HCJiang}.
Although the mean-field phase diagram contains a number of candidate QSLs, VMC
calculations reveal that only two are robust. One is the GKSL, whose regime of
stability is bounded approximately by $|J/K| = 0.2$ at $\Gamma = 0$ and
$\Gamma/|K|$ = 0.15; this result provides a quantitative statement of the
region of relevance for the considerations of Ref.~\cite{rsyb}. The second
we name the PKSL, one of our central results being that there is only one QSL
proximate to the GKSL. The GKSL and PKSL have the same PSG despite being
physically quite different states. In contrast to the spinon excitation
spectrum of the GKSL, which has two Majorana cones in the first Brillouin
zone [Fig.~\ref{Fig_Mcones}(a)], the PKSL has 14 [Figs.~\ref{Fig_Mcones}(b) and
\ref{Fig_Mcones}(d)]. These cones are protected from local perturbations by
the combination of spatial-inversion and time-reversal symmetry, as detailed
in Sec.~S3 of the SM \cite{sm}. We discuss the nature of the cones and the
magnetic response of the PKSL below.

The majority of the phase diagram (Fig.~\ref{Fig_phsdgrm}) consists of the
magnetically ordered states familiar from the classical $K$-$J$-$\Gamma$ model
\cite{c1}, namely AFM, stripe, incommensurate spiral (IS), zigzag, and FM
ordered phases. Unsurprisingly, the boundaries around the QSL phases are rather
different in the quantum model, with FM being stronger at low $\Gamma$ and IS
extending to $J < 0$. We comment that, not only can an incommensurate ordered
phase still exist for $S = 1/2$, but it can also exist throughout the phase
diagram from (near) a pure $K$-$\Gamma$ to a pure $J$-$\Gamma$ model; as
expected of our ordered phases, the spiral angles we obtain agree with
Ref.~\cite{c1}. We draw attention to the fact that all of the $J = 0$ line
in Fig.~\ref{Fig_phsdgrm} is ordered for $\Gamma/|K| > 0.55$, with IS order
until $\Gamma/|K| = 0.7$ and zigzag order all the way to and in the pure
$\Gamma$ model \cite{rliao}. This result contradicts an iDMRG study \cite{c3}
that reported only QSL phases in the $K$-$\Gamma$ model; as we show in Sec.~S4
of the SM \cite{sm}, that conclusion is relevant only for very narrow
cylinders. While the existence of order has been called into question at
$J = 0$ \cite{rjav2}, it is robust in the $S = 1/2$ model in VMC.

We state without demonstration that all of the phase transitions between the
magnetically ordered phases are first-order, which is the simplest possibility
when two phases of differing order parameters meet. A more complex situation
is possible for the transitions between ordered and QSL phases, which could
be second-order or even show an intermediate phase with coexisting magnetic
and Z$_2$ topological order (Sec.~S3 of the SM \cite{sm}). However, we find at
all points we have investigated that these transitions are also first-order.
Finally, the transition between the GKSL and PKSL is also sharply first-order,
as may be observed both in the ground-state energy [Fig.~\ref{Fig_Mcones}(c)]
and through discontinuities in the optimal variational parameters (not shown).
Neither the GKSL nor the PKSL dispersion [Figs.~\ref{Fig_Mcones}(a) and
\ref{Fig_Mcones}(b)] evolves significantly around the level-crossing.

Returning to the spin response of the QSLs, it is helpful to consider
Kitaev's representation \cite{c0} of the spin in terms of Majorana fermions,
$S^m = i b^m c$ ($m = x, y, z$), which we review in Sec.~S1 of the SM
\cite{sm}. In the GKSL, the $c$ fermions are gapless and the $b^m$ fermions
gapped, but the gapped spin response of the KSL becomes gapless when both
$J$ and $\Gamma$ are non-zero \cite{rsyb}. This is verified in our VMC
calculations in the form of $c$-$b^m$ hybridization in the low-energy limit,
but as shown in Sec.~S5 of the SM \cite{sm} this hybridization remains very
weak throughout the GKSL regime, such that low-energy spin excitations, if
present, have very little weight. By contrast, in the PKSL we find that the
quasiparticles are strongly hybridized combinations of $c$ and $b^m$ fermions
at all wave vectors and energies, which is a consequence of the finite $\eta_3$
parameter induced by the $\Gamma$ term. Thus low-energy spin excitations arise
from both intra- and intercone spinon processes and the spin response of the
PKSL is gapless, as we illustrate in Fig.~\ref{Fig_Mcones}(e) by computing the
dynamical structure factor, $S({\pmb q},\omega)$, at the mean-field level for
energy $\omega = 0$ (Sec.~S5). The positions of the maxima are readily
understood from the cone structure shown in Fig.~\ref{Fig_Mcones}(d). For
energies beyond the cone region, $S({\pmb q},\omega)$ takes on a complex and
continuous form [Fig.~\ref{Fig_Mcones}(f)].

One of the most exciting properties of the KSL is that it becomes a gapped,
non-Abelian CSL in an applied magnetic field of any orientation not orthogonal
to an Ising (spin) axis. Specifically, Kitaev classified 16 different types of
CSL based on the statistics of their vortices, which are defects arising from
inserted flux quanta \cite{c0}. He showed that these Z$_2$ vortices are Abelian
anyons when the Chern number, $\nu$, is even and non-Abelian anyons when $\nu$
is odd, leading to a clear illustration of topological properties, edge modes,
fusion rules, and applications in quantum computation. The KSL in a field
provides an example of the class $\nu = 1$, where the non-Abelian statistics
arise due to unpaired Majorana modes ($\sigma$ in Table \ref{tab:mass})
associated with the vortices. By VMC calculations in a field $\pmb B \parallel
{\hat c}$ (Sec.~S6 of the SM \cite{sm}), we verify at the mean-field level that
these modes are also present in the GKSL; this non-Abelian regime is terminated
at a Z$_2$ deconfinement-to-confinement transition \cite{HCJiang}.

\begin{figure}[t]
\includegraphics[scale=0.42]{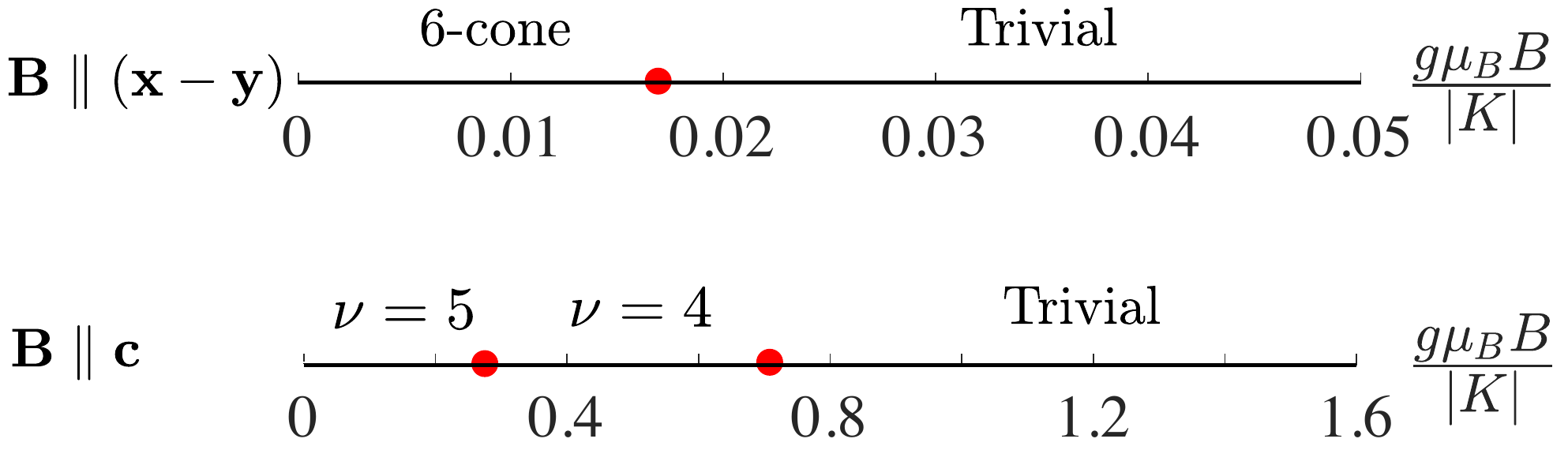}
\caption{Phase diagrams of the PKSL ($\Gamma/|K| = 0.3$, $J = 0$) in a magnetic
field applied in the ${\hat x} - {\hat y}$ direction and in the ${\hat c}
 = {\textstyle \frac{1}{\sqrt3}} (\hat x + \hat y + \hat z)$ direction.
``6-cone'' denotes a phase whose low-energy spinon dispersion has six
remaining cones.}
\label{Fig_pksl_field}
\end{figure}

In the PKSL, each of the 7 pairs of cones becomes gapped in a field
$\pmb B \parallel {\hat c}$ and contributes a Chern number $\nu = \pm
1$. We demonstrate by VMC calculations at $\Gamma/|K| = 0.3$ and $J = 0$
that a CSL phase with $\nu = 5$ is obtained at small $|\pmb B|$. As shown
in Fig.~\ref{Fig_pksl_field}, there are two successive, weakly first-order
phase transitions with increasing $|\pmb B|$, to a $\nu = 4$ CSL and then
to a trivial phase connected to the fully polarized state. (We comment that
this trivial, Z$_2$-confined phase is obtained from both the $\nu = 1$ and
$\nu = 0$ mean-field states, as shown in Sec.~S6 of the SM \cite{sm}.) Thus
the PKSL provides a specific realization of two little-known cases from
Kitaev's ``16-fold way,'' and we use it in Sec.~S6 to illustrate their vortex
modes at the mean-field level. The $\nu = 5$ phase is a non-Abelian CSL while
the $\nu = 4$ phase is Abelian. Quite generally, in a CSL with Chern number
$\nu$, there are $\nu$ branches of chiral Majorana edge states, each of which
contributes to a total chiral central charge $c_- = \nu/2$. The thermal Hall
conductance, which is a physical observable, is therefore quantized to
$\kappa_{xy} / T \Lambda = c_-$, where $\Lambda = \pi k_B^2 / 6h$ is a
constant. From the number of associated mid-gap modes it can be shown
that the vortex carries topological spin $e^{i\nu {\pi\over 8}}$, and from its
fusion with itself or with a fermion we obtain the anyonic character of the
CSLs as listed in Table \ref{tab:mass}. These results are further verified
within our VMC analysis by calculating the ground-state degeneracy (GSD) on
a torus, which matches the number of topologically distinct quasiparticle
types.

\begin{table}[t]
\centering
\begin{tabular}{c c c c c}
\hline
\hline
Parent state  & CSL & $\;$topological sectors $\;$   & $\;$GSD$\;$& $c_{-}$ \\
\hline
PKSL          & $\; \nu = 5 \;$   & $\sigma,\varepsilon,1$  & 3      & 5/2 \\
PKSL          & $\nu = 4$    &$e, m, \varepsilon,1$   & 4      & 2 \\
KSL           & $\nu = 1$   & $\sigma,\varepsilon,1$  & 3  & 1/2 \\ 
U(1) Dirac    & KL   & $e,1$                      & 2      & 1 \\ 
\hline
\hline
\end{tabular}
\caption{Field-induced CSL states realized to date in $K$-$J$-$\Gamma$ models.
$\nu$ is the Chern number. KL is the Kalmayer-Laughlin state \cite{rkl,c2}. 1
denotes the vacuum and $\varepsilon$ the fermion. $\sigma$ denotes the vortices
in the non-Abelian phases ($\nu = 1$ and 5), which have topological spin
$e^{i{\pi\over8}}$ and $e^{i{5\pi\over8}}$. $e$ and $m$ are the two different types of
vortex in the Abelian CSL ($\nu = 4$), which are both semions. GSD abbreviates
the ground-state degeneracy on a torus. $c_-$ is the chiral central charge.}
\label{tab:mass}
\end{table}

In contrast to the emergence in an arbitrary applied field of fully gapped
quantum states, which can be distinguished by their $\nu$ or edge $c_-$
numbers, is the possibility that, for specific field directions, the system
remains a gapless Z$_2$ QSL over a finite range of field magnitudes \cite{c2}.
In the PKSL we find this to be the case for $\pmb B \parallel (\pmb x - \pmb
y)$, where the six black and blue cones shown in Fig.~\ref{Fig_Mcones}(d)
remain gapless at low fields, whereas the other cones become gapped. For $\pmb
B \parallel (\pmb y - \pmb z)$, the black and red cones remain gapless and
for $\pmb B \parallel (\pmb z - \pmb x)$ the black and green cones. With
increasing field, pairs of cones move towards each other, but a first-order
transition occurs to a fully gapped state at a critical field $B_c \simeq
0.017 {|K|/ g\mu_B}$ (Fig.~\ref{Fig_pksl_field}). This reflects the fact that
many competing states appear at these low energy scales. We leave the response
of the PKSL to fields of arbitrary orientation and strength to a future study.

The $K$-$J$-$\Gamma$ model is expected to provide the basis for
interpreting the physics of 4$d$ and 5$d$ transition-metal compounds. While
robust magnetic order suggests strong $J$ terms in Li$_2$IrO$_3$ and
Na$_2$IrO$_3$ \cite{ryea,rchoietal}, $\alpha$-RuCl$_3$ is thought to have
only weak $J$ \cite{rwdyl} and many different $K$ and $\Gamma$ combinations
have been suggested \cite{rjav2,rcm}. Our phase diagram provides a
quantitative guide to the low-$J$, low-$\Gamma$ parameter regime required
to observe QSL behavior, including by application of pressure to the known
candidate materials. It also offers a detailed framework within which to
interpret both the physics of the magnetically disordered material
H$_3$LiIr$_2$O$_6$ \cite{rhliiro} and the reported observation of a
half-integer-quantized thermal Hall conductance in $\alpha$-RuCl$_3$
\cite{ThermalHall}.

In summary, we have obtained the phase diagram of the quantum $K$-$J$-$\Gamma$
model on the honeycomb lattice. We find two quantum spin-liquid phases, the
generic KSL and one proximate KSL, that have the same projective symmetry
group but quite different low-energy physics. The PKSL is a gapless Z$_2$ QSL
with 14 Majorana cones and a gapless spin response. In an applied field it
realizes a gapped, non-Abelian chiral QSL with $\nu = 5$ and an Abelian one
with $\nu = 4$. We also map the wide variety of magnetically ordered phases
appearing in the quantum limit. Our phase diagram provides an essential guide
to the physics of candidate Kitaev materials.

{\it Acknowledgments.} We thank M. Cheng, H.-J. Liao, Q. Luo, M. Vojta,
C. Wang, T. Xiang, Q. Wang, and especially Y.-M. Lu and L. Zou for helpful
discussions. This work was supported by the Ministry of Science and Technology
of China (Grant No.~2016YFA0300504), the NSF of China (Grant No.~11574392 and No.~11974421),
and the Fundamental Research Funds for the Central Universities and the
Research Funds of Renmin University of China (No.~19XNLG11).

\end{document}


\setcounter{equation}{0}
\renewcommand{\theequation}{S\arabic{equation}}
\setcounter{figure}{0}
\renewcommand{\thefigure}{S\arabic{figure}}
\setcounter{section}{0}
\renewcommand{\thesection}{S\arabic{section}}
\setcounter{table}{0}
\renewcommand{\thetable}{S\arabic{table}}

\onecolumngrid

\vskip1cm

\centerline{\large {\bf {Supplemental Material for ``One Proximate Kitaev
Spin Liquid}}}

\vskip1mm

\centerline{\large {\bf {in the $K$-$J$-$\Gamma$ Model on the Honeycomb
Lattice''}}}

\vskip4mm

\centerline{Jiucai Wang, B. Normand, and Zheng-Xin Liu}

\vskip8mm

\twocolumngrid

\section{Fermionic Spinon Representation}

\subsection{SU(2) gauge structure}\label{SU2Gauge}

A detailed introduction to the fermionic spin representation, by which the
spinons $C = (c_\up, c_\dn)^T$ are used to express the spin operators as $S^m
 = {\textstyle \frac12} C^\dag \sigma^m C$, may be found in Sec.~S1 of
Ref.~\cite{c2}. The two spinon species may further be expressed in terms
of four Majorana fermions,
\begin{equation}
c_\up = {\textstyle \frac12} (b^z + ic),\ \ \
c_\dn = {\textstyle \frac12} (b^x + ib^y),
\label{emfd}
\end{equation}
which satisfy the anti-commutation relations $\{b^\alpha, b^\beta\} = 2
\delta^{\alpha \beta}$ ($\alpha,\beta = 0,x,y,z$; $b^0 \equiv c$). In this
basis, the spin operator takes the form
\begin{equation}
S^m = ib^mc,
\end{equation}
and the particle-number constraint, $C^\dag C = 1$, is $b^x b^y b^z c = 1$.

The anti-commutation relation of the Majorana fermions is invariant under the
group SO(4). The relation SO(4) $\simeq$ SU(2)$\times$SU(2)/Z$_2$ expresses
the fact that the Majorana fermions are invariant under two types of SU(2)
operation, one being the expected spin-rotation group and the other a
``particle-hole-type'' charge rotation, which behaves as an internal gauge
symmetry for the fermionic representation of spin $S = 1/2$.

It is convenient to introduce the time-reversal and particle-hole partner of $C$, namely $\bar C
 = (c_\dn^\dag , -c_\up^\dag)^T$, in terms of which the spin operator may also
be expressed as $S^m = {\textstyle \frac12} \bar C^\dag \sigma^m \bar C$. $C$
and $\bar C$ can be combined into a matrix operator,
\begin{equation*}
\psi = \Bmat C & \bar C \Emat = \Bmat c_\up &  c_\dn^\dag \\ c_\dn & -c_\up^\dag
 \Emat = {\textstyle \frac12} (ic + b^x \tau_x + b^y \tau_y + b^z
\tau_z),
\end{equation*}
to obtain a symmetric representation of the spin operators as
$S^m = {\textstyle \frac{1}{4}} {\rm Tr}(\psi^{\dagger} \sigma^m \psi)$,
while the particle-number constraint is ${\rm Tr}(\psi \tau^z \psi^\dag) = 0$,
where we use the Pauli matrix $\tau^z$ to distinguish from the spin operation
$\sigma^z$.

In this framework, a left SU(2) transformation, $\psi \to g^\dag \psi$ (or
equivalently $\psi^\dag \to \psi^\dag g$), mixes $c_\up$ and $c_\dn$ and thus
corresponds to a spin rotation,
\[{\textstyle \frac{1}{4}} {\rm Tr} (\psi^{\dagger}g \sigma^m g^\dag \psi)
= \sum_{n} R_{nm} (g) S^{n}, \]
where $R(g)$ is an SO(3) matrix representation of $g$. By contrast, a right
SU(2) transformation, $\psi \to \psi W$, mixes $C$ and $\bar C$ and
corresponds to the SU(2) gauge symmetry,
\[{\textstyle \frac{1}{4}} {\rm Tr} (W^\dag \psi \tau^m \psi^{\dagger} W) =
\sum_n R (W)_{nm} \Lambda^{n},\]
where $\Lambda^m = {\textstyle \frac{1}{4}} {\rm Tr} (\psi \tau^m \psi^{\dag})$
are the generators of the SU(2) gauge transformation in the same way that the
spin operators $S^{m} = {\textstyle \frac{1}{4}} {\rm Tr} (\psi^{\dag} \sigma^m
\psi)$ are the generators of the SU(2) spin rotation. It is easy to see that
$\pmb \Lambda$ is spin-rotation-invariant and $\pmb S$ is SU(2) gauge-invariant.
In this notation, the particle-number constraint (${\rm Tr} (\psi \tau^z
\psi^\dag) = 0$) can be written in the SU(2) gauge-covariant form $\pmb
\Lambda = 0$, i.e.
\beq\label{cstrn}
{\rm Tr} (\psi {\pmb \tau} \psi^{\dagger}) = 0,
\eeq
which specifies that the physical Hilbert space is SU(2) gauge-invariant.

\subsection{Mean-field Hamiltonian}\label{s2}

The Heisenberg exchange interaction can be written as
\beq
\pmb {S}_{i} \! \cdot \! \pmb {S}_{j} & = & - {\textstyle \frac{1}{8}} {\rm Tr}
\, (\psi_{j}^{\dag} \psi_{i} \psi_{i}^{\dagger}\psi_{j}) \label{singlet_DC} \\ & = &
{\textstyle \frac{1}{8}} {\rm Tr} \, (\psi_{j}^{\dagger} \pmb{\sigma} \psi_{i} \!
\cdot \! \psi_{i}^\dag \pmb{\sigma} \psi_{j}) \nonumber
\eeq
up to constant terms. To make more transparent the connection to the spinon
representation, we note that the singlet matrix operator, $\psi_{i}^{\dagger}
\psi_{j} $, and the triplet operator, $\psi_{i}^{\dagger} \pmb \sigma \psi_{j}$,
can be expanded as
\begin{eqnarray*}
\psi_{i}^{\dagger} \psi_{j} = \left( \begin{matrix} C_{i}^{\dagger} C_{j} &
C_{i}^{\dagger}\bar{C}_{j} \\ \bar{C}_{i}^{\dagger} C_{j} & \bar{C}_{i}^{\dagger}
\bar{C}_{j} \end{matrix} \right) \!,\ \ \psi_{i}^{\dagger} \pmb{\sigma} \psi_{j} =
\left( \begin{matrix} C_{i}^{\dagger} \pmb{\sigma} C_{j} & C_{i}^{\dagger}
\pmb{\sigma} \bar{C}_{j} \\ \bar{C}_{i}^{\dagger} \pmb{\sigma} C_{j} &
\bar{C}_{i}^{\dagger} \pmb{\sigma} \bar{C}_{j} \end{matrix} \right) \! .
\end{eqnarray*}
The sign difference between the two decoupling formulas in
Eq.~(\ref{singlet_DC}) means that the version in the first line is suitable
for AFM Heisenberg interactions and that in the second line for FM ones.

In the mean-field approximation, one decouples these interactions to obtain a
non-interacting fermionic Hamiltonian. In the AFM case ($J > 0$)
\beq
H_{\rm mf} = - J \sum_{\langle i,j\rangle} {\rm Tr} \big( U_{ij}^\dag \psi_i^\dag \psi_j
 + {\rm h.c.} \big) + \sum_i \pmb \lambda \! \cdot \! \pmb \Lambda_i,
\label{MF_singlet}
\eeq
while in the FM one ($J < 0$)
\beq
H_{\rm mf} = J \sum_{\langle i,j\rangle} {\rm Tr} \big( \pmb U_{ij}^\dag \! \cdot
\! \psi_i^\dag \pmb \sigma \psi_j + {\rm h.c.} \big) + \sum_i \pmb \lambda
\! \cdot \! \pmb\Lambda_i, \label{MF_triplet}
\eeq
where $\pmb \lambda$ are the Lagrange multipliers corresponding to
Eq.~(\ref{cstrn}), $U_{ij}^\dag = U_{ji} \propto \langle \psi_j^\dag \psi_i
\rangle$ is a matrix of singlet parameters and $\pmb U_{ij}^\dag = \pmb U_{ji}
\propto \langle \psi_j^\dag \pmb \sigma\psi_i\rangle$ is a triplet parameter
matrix. Under the gauge transformation $\psi_i \to \psi_i W_i$,
\beq
{\rm Tr} \big( U_{ij}^\dag \psi_i^\dag \psi_j  \big) \to \; && {\rm Tr} \, \big(
U_{ij}^\dag W_i^\dag \psi_i^\dag \psi_j W_j \big) \nonumber \\ && = {\rm Tr}
\big( W_j U_{ij}^\dag W_i^\dag \psi_i^\dag \psi_j \big),
\eeq
meaning that $U_{ji} \to U_{ji}' = W_j U_{ji} W_i^\dag$ varies as an SU(2) gauge
field and similarly for $\pmb U_{ji} \to \pmb U_{ji}' = W_j \pmb U_{ji} W_i^\dag$.
Thus the spin system is described in the mean-field theory by free fermions
coupling to static SU(2) gauge fields on the bonds of the honeycomb lattice.

In insulating systems with spin-orbit coupling, the spin interactions
are anisotropic, as expressed in the $K$-$J$-$\Gamma$ Hamiltonian
[Eq.~(1) of the main text]. In this case the mean-field Hamiltonian
will contain both singlet ($\psi_{i}^{\dagger} \psi_{j}$) and triplet terms
($\psi_{i}^{\dag} \pmb{\sigma} \psi_{j} $). As an example, an AFM Ising
interaction can be decoupled \cite{c2} as
\beq
S^m_i S^m_j = {\textstyle -\frac{1}{16}} \Big[ {\rm Tr} (\psi_i^\dag \psi_j
\psi_j^\dag \psi_i) + {\rm Tr} (\psi_i^\dag \sigma^m \psi_j \psi_j^\dag \sigma^m
\psi_i) \big], \nonumber
\nonumber \label{Ising}
\eeq
up to a constant term, where $m = x,y,z$, and thus any XXZ-type interaction
may be expressed by adding triplet hopping and pairing terms of the form
$\sum_{\langle i,j\rangle} {\rm Tr} \big( U_{ij}^z \psi_i^\dag \sigma^z \psi_j \big)$
to the mean-field Hamiltonian of Eq.~(\ref{MF_singlet}). The most general
mean-field Hamiltonian for a spin-liquid state in a spin-orbit-coupled
magnetic insulator, containing both (\ref{MF_singlet}) and (\ref{MF_triplet}),
takes the form
\beq
H_{\rm mf}^{\rm SL} = \sum_{\langle i,j\rangle} \! {\rm Tr} \big( U_{ji} \psi_i^\dag \psi_j
 + \pmb U_{ji} \! \cdot \! \psi_i^\dag \pmb \sigma \psi_j  + {\rm h.c.} \big)
 + \sum_i \! \pmb \lambda \! \cdot \! \pmb \Lambda_i, \nonumber \\
\label{gmf}
\eeq
where $U_{ji}$ and $\pmb U_{ji}$ are treated as variational parameters.

\section{Invariant Gauge Group and Projective Symmetry Group}\label{IGGPSG}

Because of the static bond gauge fields contained in $U_{ji}$ and $\pmb U_{ji}$,
the mean-field Hamiltonian of Eq.~(\ref{gmf}) is not in general invariant under
an arbitrary SU(2) gauge transformation. The subgroup of the SU(2) gauge group
under which Eq.~(\ref{gmf}) remains invariant is called the invariant gauge
group (IGG) of the spin-liquid state. The IGG determines the nature of the
low-energy gauge fluctuations in a QSL phase.

If the IGG is U(1), there exists an SU(2) gauge transformation that eliminates
all fermion-pairing terms in the mean-field Hamiltonian, meaning that the
number of fermionic slave particles is conserved. In this case, the gapless
gauge photons contribute to the low-energy excitations. In two spatial
dimensions (2D), it is known that a U(1) gauge field is confined unless it
is coupled to a matter field (in this case the fermions) that is either (i)
gapless or (ii) has a nonzero Chern number.

If the IGG is Z$_2$, as in the KSL, fermion-pairing terms cannot be removed by
any SU(2) gauge transformation. The Z$_2$ gauge-flux excitations in this case
are usually gapped, and can remain deconfined even if the matter field is
gapped with zero Chern number. The low-energy gauge fluctuations dictated by
the confinement or deconfinement of the Z$_2$ gauge field are reflected in
the ground-state degeneracy (GSD) of the QSL when placed on a torus: in the
thermodynamic limit, inserting a global Z$_2$ $\pi$ flux in either hole of
the torus costs no energy. Because this process is equivalent to exchanging
the boundary conditions of the mean-field Hamiltonian from periodic to
anti-periodic, in 2D one may construct the four mean-field ground states
$|\psi_{\pm\pm} \rangle$, where the subscripts denote the boundary conditions
for the $x$- and $y$-directions. After a Gutzwiller projection of these four
states to the physical Hilbert space, the number of linearly independent
states is equal to the GSD on a torus.

We calculate the density matrix of the projected (VMC) states from the
wave-function overlap $\rho_{\alpha\beta} = \langle P_G \psi_\alpha | P_G \psi_\beta
\rangle = \rho_{\beta\alpha}^*$, with $\alpha,\beta \in \{++,+-,-+,--\}$.
If $\rho_{\alpha\beta}$ has only one significant eigenvalue, with the others
vanishing, then the GSD is 1, indicating that the Z$_2$ gauge field is
confined. If $\rho_{\alpha\beta}$ has more than one near-degenerate nonzero
eigenvalue, the GSD is nontrivial and hence the Z$_2$ gauge fluctuations
are deconfined. In this deconfined phase, if the Chern number is even then
from above the GSD is 4; however, if the Chern number is odd, as we will
find in Sec.~\ref{SLinField}, then the GSD is 3 because the mean-field
ground state $|\psi_{++} \rangle$ has odd fermionic parity and vanishes
after Gutzwiller projection.

Quite generally, a QSL should respect all the symmetries of the spin
Hamiltonian, but at the mean-field level this constraint can be relaxed
in the following sense. Under a symmetry operation $g$, $H_{\rm mf}$ may be
transformed to a different expression, $g H_{\rm mf}  = H_{\rm mf}' \neq H_{\rm mf}$,
but if it can be transformed back to its original form by an SU(2) gauge
transformation then this mean-field Hamiltonian still describes a spin-liquid
state. Specifically,
\beq
\psi_i & \to & g^\dag \psi_{g(i)} W_i(g), \label{psipsg} \\
{\rm Tr} \, [\pmb U_{ji} \! \cdot \! \psi_i^\dag\pmb\sigma\psi_j] & \to & {\rm
Tr} \, [W_{j} \pmb U_{ji} W_{i}^\dag \! \cdot \! \psi_{g(i)}^\dag g \pmb \sigma g^\dag
\psi_{g(j)}] \nonumber \\ & = & {\rm Tr} \, [\pmb U_{g(j)g(i)} \! \cdot \!
\psi_{g(i)}^\dag \pmb \sigma \psi_{g(j)}], \label{psipsg2} \\
{\rm Tr} \, [U_{ji} \psi_i^\dag \psi_j] & \to & {\rm Tr} \, [W_{j} U_{ji} W_{i}^\dag
\psi_{g(i)}^\dag g g^\dag \psi_{g(j)}] \nonumber \\ & = & {\rm Tr} \, [U_{g(j)g(i)}
\psi_{g(i)}^\dag \psi_{g(j)}], \label{psipsg3}
\eeq
which requires $U^n_{g(j)g(i)} = \sum_m R_{nm}(g) W_{j}(g) U^m_{ji} W_{i}^\dag(g)$
and $U_{g(j)g(i)} = W_{j}(g) U_{ji} W_{i}^\dag(g)$. The new symmetry operations,
each of which involves a symmetry operation $g$ followed by a gauge
transformation $W_i(g)$, form a larger group that is known as the projective
symmetry group (PSG). Together with the IGG, the PSG of the mean-field
Hamiltonian is used to classify and distinguish the different spin-liquid
phases.

The gapless KSL is believed \cite{rsyb} to be a finite, stable phase in the
presence of non-Kitaev interactions, including the $J$ and $\Gamma$ terms.
The mean-field Hamiltonian describing the generic state around the KSL, which
we denote the gKSL, will then respect the same PSG as the KSL itself. Besides translation symmetry, the
symmetry group of the pure KSL, $G = D_{3d}$$\times$$Z_2^T$, has the three
generators
\[
S_6 = (C_3)^2 P,\ \ \  M = C_2^{x-y} P,\ \ \   T = i\sigma^y K,
\]
where $C_3$ is a threefold rotation around the direction $\hat c = {\textstyle
\frac{1}{\sqrt3}} (\hat x + \hat y + \hat z)$, $C_2^{x-y}$ is a twofold rotation
around ${\textstyle \frac{1}{\sqrt2}} (\hat x-\hat y)$, and $P$ is spatial
inversion. The PSG of the KSL is read most simply from the Majorana
representation, in which the mean-field Hamiltonian is
\begin{eqnarray}
H_{\rm mf}^{K} & = & \sum_{\langle i,j \rangle\in\gamma} \rho_a (ic_ic_j) +
\rho_c (i b_i^\gamma b_j^\gamma) \nonumber \\ & = & \sum_{\langle i,j \rangle\in\gamma}
i \rho_a {\rm Tr} (\psi_i^\dagger \psi_j + \tau^x \psi_i^\dagger
\sigma^x \psi_j + \tau^y \psi_i^\dagger \sigma^y \psi_j \nonumber \\ & &
\;\;\;\;\;\;\; + \tau^z \psi_i^\dagger \sigma^z \psi_j) + i \rho_c {\rm Tr}
(\psi_i^\dagger \psi_j + \tau^\gamma \psi_i^\dagger \sigma^\gamma \psi_j \nonumber
\\ & & \;\;\;\;\;\; - \tau^\alpha \psi_i^\dagger \sigma^\alpha \psi_j - \tau^\beta
\psi_i^\dagger \sigma^\beta \psi_j) + {\rm h.c.}
\label{Kitaevmf}
\end{eqnarray}
Because the $c$ fermion never mixes with any of the $b^m$ fermions, any PSG
operation leaves the $c$ fermions invariant. The gauge operation, $W_i(g)$,
following the symmetry operation $g$ should then be $W_i(g) = \pm g$. A
detailed analysis \cite{c5} shows that the gauge transformations of the
generators $S_6$, $M$, and $T$ are
\beq\label{PSGKSL}
W_A(S_6) & =& - W_B(S_6) = \exp \! \big[-i {\textstyle \frac{4}{3}} \pi
{\textstyle \frac{1}{2\sqrt{3}}} (\tau^x + \tau^y + \tau^z) \big],
\nonumber \\
W_A(M) & =& - W_B(M) = \exp \! \big[-i \pi {\textstyle \frac{1}{2\sqrt{2}}}
(\tau^x - \tau^y) \big], \nonumber\\
W_A(T) & = &- W_B(T) = i \tau^y,
\eeq
where A and B denote the two sublattices of the honeycomb lattice.

\section{$K$-$J$-$\Gamma$ Model: Symmetry-Protection and Magnetic Order}
\label{s4}

When the Kitaev model is extended to the $K$-$J$-$\Gamma$ model, still with
only nearest-neighbor coupling on the honeycomb lattice for all three types
of term, there are several different ansatzes for states beyond the Kitaev
mean-field Hamiltonian [Eq.~(\ref{Kitaevmf})] that are invariant under the
same PSG. The $\Gamma$ interaction gives rise to the mean-field terms
\beq\label{Gm}
H_{\rm mf}^{\Gamma} & = & \!\! \sum_{\langle i,j\rangle\in\gamma} \!\! i\rho_d
(b_i^\alpha b_j^\beta + b_i^\beta b_j^\alpha) \\ & = & \!\! \sum_{\langle i,j\rangle
\in \gamma} \!\! i\rho_d {\rm Tr} \left( \tau^\alpha \psi_i^\dag
\sigma^\beta \psi_j + \tau^\beta \psi_i^\dag \sigma^\alpha \psi_j\right)
 + {\rm h.c.} \nonumber
\eeq
It is readily demonstrated from the Majorana decoupling that Heisenberg
interactions do not give rise to any new types of term. Thus when expressed
in terms of $U_{ij}$ and $\pmb U_{ij}$, the full expression of the mean-field
Hamiltonian that preserves the PSG is
\beq\label{MFPSG}
H_{\rm mf}^{\rm SL} & = & \!\!\!\! \sum_{\langle i,j \rangle\in\gamma} \!\!\! {\rm Tr}
\, [U_{ji}^{(0)} \psi_i^\dag \psi_j] \! + \! {\rm Tr} \, [U_{ji}^{(1)} \psi_i^\dag
(i R_{\alpha\beta}^\gamma) \psi_j] \\ & & \;\;\;\; + {\rm Tr} \, [U_{ji}^{(2)}
\psi_i^\dag \sigma^\gamma \psi_j] \! + \! {\rm Tr} \, [U_{ji}^{(3)} \psi_i^\dag
\sigma^\gamma R_{\alpha\beta}^\gamma \psi_j] \! + \! {\rm h.c.}, \nonumber
\eeq
where $R_{\alpha\beta}^\gamma = - \frac{i}{\sqrt{2}} (\sigma^\alpha + \sigma^\beta)$,
which is Eq.~(2) of the main text. The decouplings expressed in
Eqs.~(\ref{Kitaevmf}) and (\ref{Gm}) contribute the terms
\begin{equation}
\begin{aligned}
& {\tilde U}_{ji}^{(0)} = i (\rho_a + \rho_c), \\
& {\tilde U}_{ji}^{(1)} = i (\rho_a - \rho_c  + \rho_d) (\tau^\alpha + \tau^\beta),
\\ & {\tilde U}_{ji}^{(2)} = i (\rho_a + \rho_c) \tau^\gamma, \\
& {\tilde U}_{ji}^{(3)} = i (\rho_c - \rho_a - \rho_d) (\tau^\alpha - \tau^\beta),
\end{aligned}
\end{equation}
to the coefficients $U_{ji}^{(m)}$, in which $j$ and $i$ specify $\gamma$.
However, the most general coefficients preserving the $C_3$ rotation symmetry
(in the PSG sense) also contain multiples of the uniform ($I$) and $\tau^x +
\tau^y + \tau^z$ gauge components,
\begin{equation}
\begin{aligned}
& {\tilde {\tilde U}_{ji}^{(0)}} = i \eta_0 + \eta_1 (\tau^x + \tau^y + \tau^z), \\
& {\tilde {\tilde U}_{ji}^{(1)}} = \eta_2 + i \eta_3 (\tau^x + \tau^y + \tau^z), \\
& {\tilde {\tilde U}_{ji}^{(2)}} = \eta_4 + i \eta_5 (\tau^x + \tau^y + \tau^z), \\
& {\tilde {\tilde U}_{ji}^{(3)}} = \eta_6 + i \eta_7 (\tau^x + \tau^y + \tau^z).
\end{aligned}
\end{equation}
If the full symmetry group, $G = D_{3d} \times Z_2^T$, is preserved, then
only the three parameters $\eta_0$, $\eta_3$, and $\eta_5$ are allowed; by
contrast, if one allows the breaking of spatial inversion symmetry, while
still preserving mirror reflection symmetry, then $\eta_1$, $\eta_2$, and
$\eta_4$ are also allowed. Thus a spin-liquid ansatz that preserves the
full PSG symmetry generated by Eq.~(\ref{PSGKSL}) contains the
variables $U_{ji}^{(m)} = {\tilde U}_{ji}^{(m)} + {\tilde {\tilde U}_{ji}^{(m)}}$
of the main text with six real scalar parameters, $\rho_a$, $\rho_c$,
$\rho_d$, $\eta_0$, $\eta_3$, and $\eta_5$. Of these, only $\eta_3$ and
$\eta_5$ lead to a hybridization of the $c$ with the $b^m$ fermions, which
as we show in Sec.~\ref{DSF} has important consequences for the spin response
of the ground state.

\subsection{Majorana Cones and Symmetry Protection}

The $G = D_{3d} \times Z_2^T$ symmetry imposes a constraint on the number of
cones in the mean-field dispersion. First, the momentum points K and K$'$ are
invariant under the subgroup $C_{3v}$, together with the combined symmetry
operation $PT$, and are transformed into each other either by time reversal
($T$) or by spatial inversion ($P$). The cones at K and K$'$ must therefore
appear as a pair. Second, if an extra cone appears at a general point $\pmb k$,
which in the present case must be invariant under one of the mirror-reflection
operations in $D_{3d}$, then the $C_3$ and $C_3^2$ operations transform $\pmb k$
into $\pmb k'$ and $\pmb k''$, respectively, while the time-reversal symmetry
further transforms these three points into $-\pmb k, -\pmb k', -\pmb k''$.
Because the spin liquid preserves the $G = D_{3d}\times Z_2^T$ symmetry, cones
should appear at all six of these points, and the total number of cones should
be $N = 2 + 6 m$, where $m \geq 0$ is an integer.

Next we discuss the symmetry protecting these cones and hence guaranteeing
their existence. It is known that the Berry phase along a small contour
surrounding a cone is quantized to $\pi$ and hence is nontrivial. However,
without symmetry-protection, the cone can become gapped and the corresponding
Berry phase can be changed continuously. The key symmetry in the present case
is the fact that the position of the cone is invariant under the combination
$PT$: for any system with $PT$ symmetry, the 1D subsystem defined by the
contour also has this symmetry. From classification theory \cite{Kitaev09,
Wen12}, a gapped 1D fermionic system preserving this combined symmetry,
with $(PT)^2 = 1$, is classified by Z$_2$, indicating that there is one
nontrivial phase. Thus the $\pi$-quantization of the Berry phase along such
a contour is $PT$-protected, and hence the stability of the cone is also
protected. The Z$_2$ classification also indicates that one mechanism for
gapping of the cones in a system preserving $PT$ symmetry is when two cones
merge.

Besides being used to construct the mean-field ansatz, the PSG can also
be used to classify QSL phases that have the same pattern of many-body
entanglement but different types of symmetry fractionalization. However,
the gKSL and PKSL phases of the present work have the same PSG but different
patterns of many-body entanglement, and hence quite different physical
properties. This difference becomes clear when a magnetic field is applied
in the $\hat c$-direction, and the resulting gapped chiral spin-liquid (CSL)
phases have different types of topological order, as we discuss in
Sec.~\ref{SLinField}.

\subsection{Magnetic Order}

In addition to spin-liquid phases, to describe the magnetic order of the
spin-symmetry-breaking phases of the $K$-$J$-$\Gamma$ model we introduce a
single-${\pmb Q}$ field \cite{c1}
\[\pmb{M}_i = M \{\sin \phi [\hat {\pmb e}_x' \cos (\pmb{Q} \cdot
\pmb{r}_i) + \hat{\pmb e}_y' \sin(\pmb{Q} \cdot \pmb{r}_i)] + \cos \phi
\, \hat{\pmb e}_z' \},\]
where the local spin axes, $\hat {\pmb e}_\alpha'$, are specified at $\pmb{r}_i
 = 0$ by three Euler angles, ($\psi,\theta,\varphi$), and $\phi$ is a canting
angle. We determine ($\psi,\theta,\varphi$) from the classical ground state.
Although $\phi$ is treated as a variational parameter, we find in all our
ordered ground states except for the IS phase that the canting angle $\phi
\to 0$. Thus the quantum corrections we compute are in essence contained in
the amplitude of the ordered moment, $M$. In this way, the static order is
treated as an external field, obtained by summing over all neighbors of a
given site, $i$, and hence the full mean-field Hamiltonian for the
$K$-$J$-$\Gamma$ model is
\begin{equation}
H_{\rm mf}^{\rm total} = H_{\rm mf}^{\rm SL} - {\textstyle \frac{1}{2}} \sum_i
(\pmb {M}_i \cdot C_i^\dagger \pmb \sigma C_i + {\rm h.c.}).
\end{equation}
The ground state of $H_{\rm mf}^{\rm total}$ preserves all of the symmetries of
the PSG or the ordered magnetic state. The trial wave functions obtained by
Gutzwiller projection then also preserve the full symmetry of the system.

\begin{figure}[t]
\includegraphics[width=8cm]{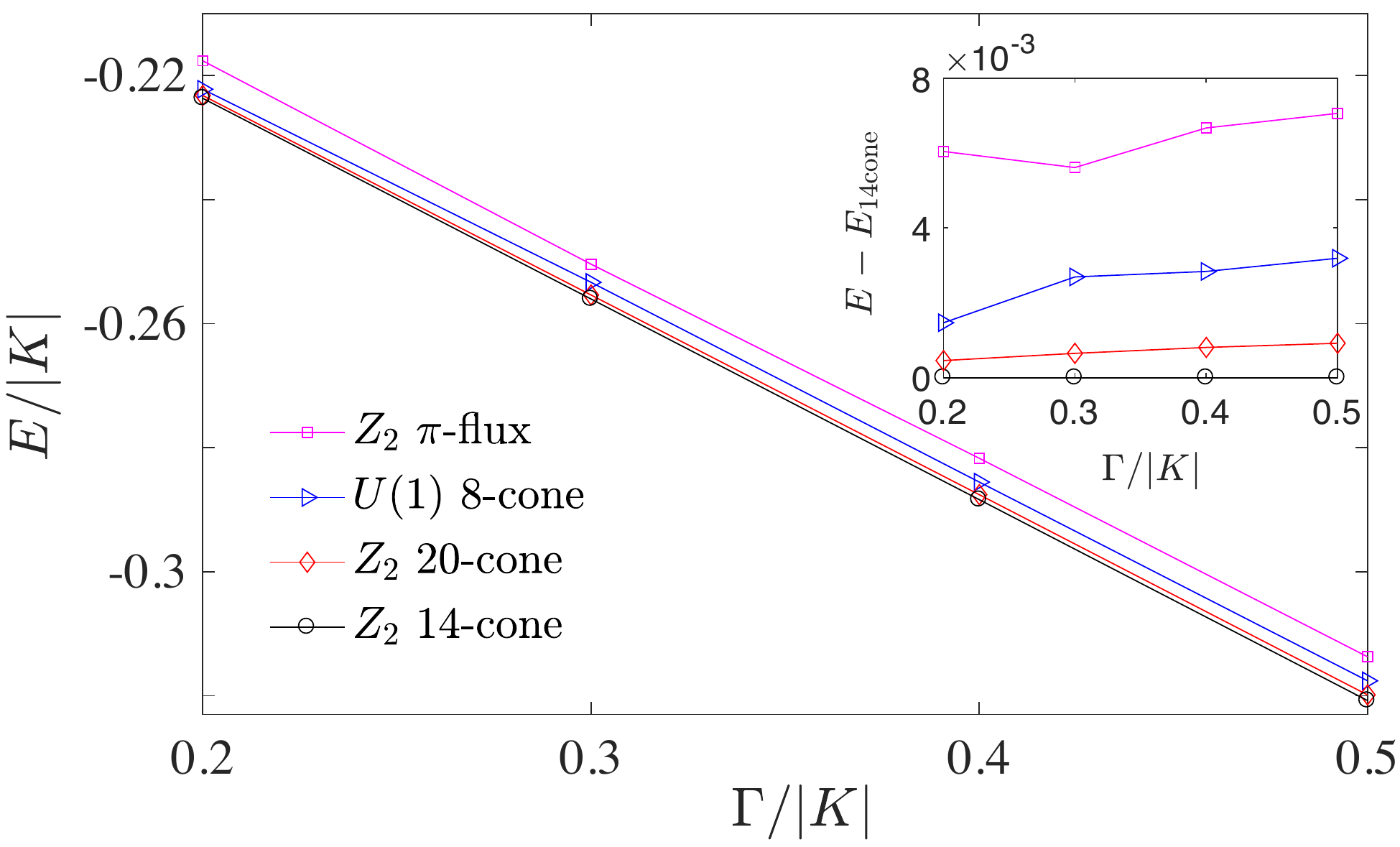}
\caption{Ground-state energy per site, $E$, of the $K$-$\Gamma$ model ($J
 = 0$), comparing the lowest-lying trial wave functions for U(1) and Z$_2$
QSLs. The inset shows the energy difference from the 14-cone Z$_2$ state.}
\label{Z2toU1}
\end{figure}

In order to provide a complete and systematic study of the $K$-$J$-$\Gamma$
phase diagram, we have investigated many different ansatzes for both the
spin-liquid and the ordered magnetic phases. In the spin-liquid regime of
Fig.~1 of the main text, one of the most important questions \cite{c2} is
whether states with IGG U(1) can compete with states whose IGG is Z$_2$
(but whose PSG may be different from that of the KSL). As Fig.~\ref{Z2toU1}
illustrates for the $J = 0$ line, in the absence of an applied field even
the leading (8-cone) U(1) state is not favored energetically in the parameter
range of the PKSL.

Turning to Z$_2$ QSLs, we have found a 20-cone state that lies very low in
energy and has the same PSG as the 14-cone state. As a technical comment,
we count the number of cones in any state by using the parameters determined
from VMC to draw the spinon dispersion of Eq.~(\ref{MFPSG}) in a momentum
space of 120$\times$120 $\pmb k$-points. Physically, although we will show
in Sec.~\ref{FiniteSize} that the 20-cone state is not favored energetically
for any large system sizes, it is instructive to present a brief discussion of
its properties. In the 20-cone state, the $b_x$, $b_y$, and $b_z$ fermions are
mutually hybridized, but in contrast to the 14-cone state they have almost no
hybridization with the $c$ fermions. This is a consequence of the variational
parameters $\eta_3$ and $\eta_5$ being finite in the 14-cone state but very
small in the 20-cone state. Further, a weak out-of-plane field applied to the
20-cone state induces an Abelian CSL with Chern number $\nu = 4$, rather than
the non-Abelian CSL with $\nu = 5$ induced from the 14-cone phase (main text).
Another candidate QSL is the $\pi$-flux state, which contains a $\pi$-flux in
every hexagon and thus has a different PSG. In summary, although the 14-cone
state is the ground state over the full PKSL regime, it is clear from
Fig.~\ref{Z2toU1} that indeed there exist several competing states which are
very close in energy, despite their quite different physical nature.

We also tested many different ansatzes for the ordered magnetic states.
Here we found without exception that ordered states with IGG Z$_2$ in
$H_{\rm mf}^{\rm SL}$ are favored. Because the gauge-flux excitations (visons)
are gapped in a Z$_2$ QSL, they may remain deconfined even in the presence
of magnetic order. Alternatively stated, a phase with coexisting magnetic
order and Z$_2$ topological order is allowed in principle, and would be
separated both from the nearest pure spin-liquid phase and from the pure
magnetically ordered phase by continuous transitions. However, our VMC
results indicate that such coexisting states do not appear in the phase
diagram: all of the magnetically ordered ground states we find have singly
degenerate ground states on a torus, indicating the absence of Z$_2$
topological order. Thus it is no surprise that all the transitions from
spin-liquid to magnetically ordered phases in Fig.~1 of the main text are
first-order.

Similar conditions are found in the presence of an applied magnetic field
(Sec.~\ref{SLinField}). Because the IGG is Z$_2$, when all the cones are
gapped by the magnetic field and the total Chern number vanishes, it remains
possible that the system can possess Z$_2$ topological order. However, such
states never appear in the VMC phase diagram, i.e.~gapped states in the
$K$-$J$-$\Gamma$ model with zero Chern number are always Z$_2$-confined and
trivial.

\section{Finite-Size Effects in the $K$-$\Gamma$ Model}\label{FiniteSize}

\begin{figure}[t]
\includegraphics[width=8.cm]{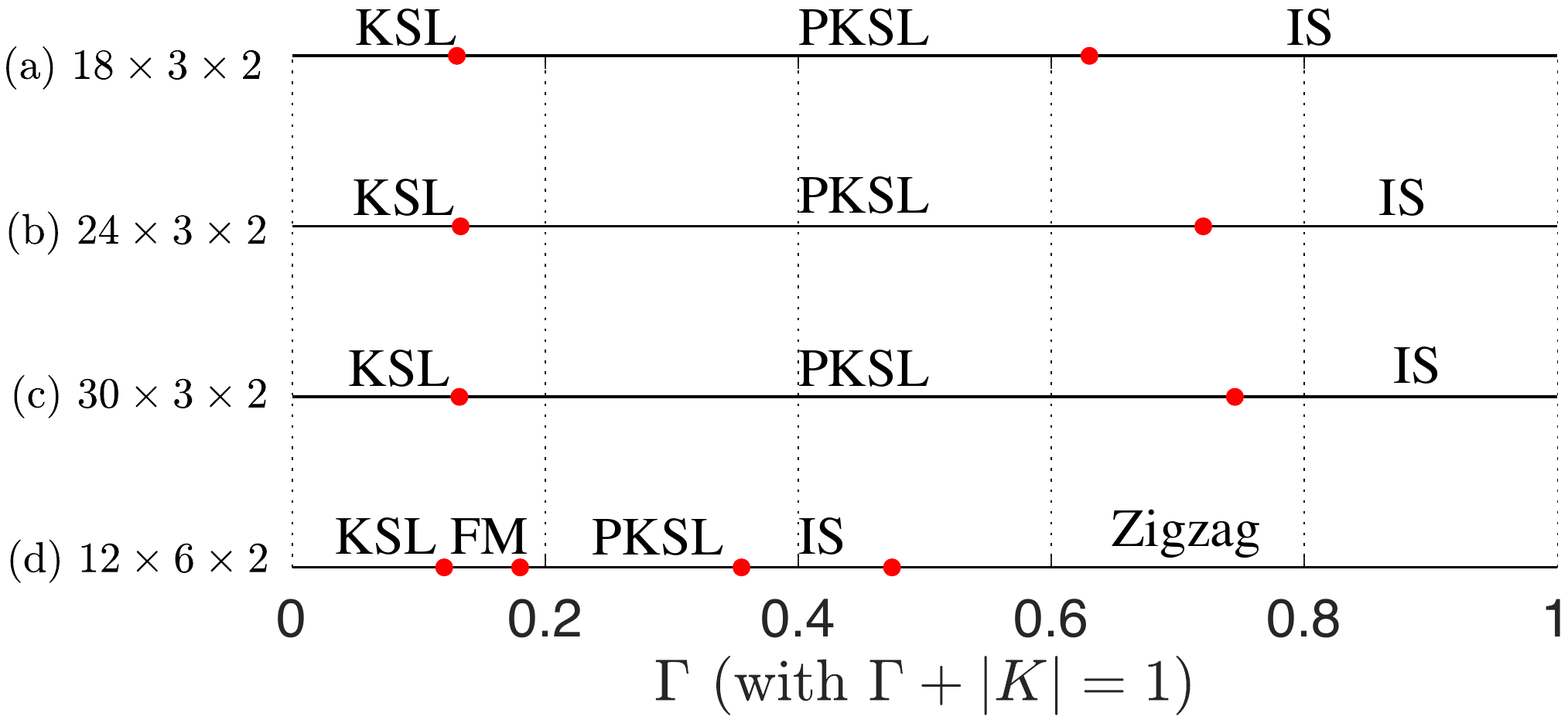}
\caption{Phase diagram of the $K$-$\Gamma$ model ($K < 0$) in a cylindrical
geometry, shown on a scale where $|K| + \Gamma = 1$. (a)-(c) Cylinders of
radius 6 sites: QSL phases dominate, their regime of stability increasing
with cylinder length. (d) System size 12$\times$6$\times$2: three magnetically
ordered phases appear. The variational parameters of the QSL phases we denote
by PKSL in panels (a)-(c) correspond in a 2D system to the 14-cone state, but
in panel (d) to the 20-cone state.}
\label{fs}
\end{figure}

As stated in the main text, the phase boundaries we find by VMC for the $K$-$J$
model, meaning the model with $\Gamma = 0$ and $J > 0$, agree quantitatively
with the results obtained by DMRG \cite{HCJ}. Specifically, we find that
$J_{c1} = 0.16|K|$ for the transition from the KSL into the spiral phase and
$J_{c2} = 0.77|K|$ for the AFM transition, while in Ref.~\cite{HCJ} $J_{c1} =
0.13|K|$ and $J_{c2} = 0.75|K|$. This indicates that finite-size effects in
the $K$-$J$ model, which in DMRG arise from the circumference of the cylinder
used, are relatively weak.

However, the situation is different for the $K$-$\Gamma$ model.
In Ref.~\cite{c3} it was stated that the ground state of the $K$-$\Gamma$
model with $K < 0$ and $\Gamma > 0$ is magnetically disordered for all values
of $\Gamma/|K|$, a result quite different from our conclusion (Fig.~1 of the
main text). We note that the iDMRG calculations of these authors were
performed on cylinders with a circumference of six lattice sites. To
reproduce this geometry, we perform VMC calculations for the model with
$J = 0$ on lattices of sizes $L_x$$\times$$L_y$$\times$2, where the
cylinder circumference is fixed to $L_y = 3$ unit cells while its length is
$L_x = 18$, 24, and 30. As shown in Figs.~\ref{fs}(a), \ref{fs}(b), and
\ref{fs}(c), indeed we find a very broad regime of QSL phases whose width
increases with $L_x$, limiting the ordered (IS) phase to values only beyond
$\Gamma/|K| = 3$ on long cylinders.

When the circumference is increased to $L_y = 6$ (system size
12$\times$6$\times$2 sites), we find a zigzag-ordered phase for all parameter
ratios $\Gamma/|K| > 0.9$, which is absent when $L_y = 3$ [Fig.~\ref{fs}(d)].
Further, a ferromagnetic phase appears at small $\Gamma$, separating the KSL
from the analog of the PKSL for this geometry (because of the finite number
of $\pmb k$ values in the transverse direction, this magnetically disordered
state will not have the full complement of gapless cones). This result alone
demonstrates that the $K$-$\Gamma$ model on the honeycomb lattice is subject to
very strong finite-size effects, and that conclusions in favor of QSL states
at large $\Gamma$ are artifacts of the system sizes considered.

\begin{figure}[t]
\includegraphics[width=7. cm]{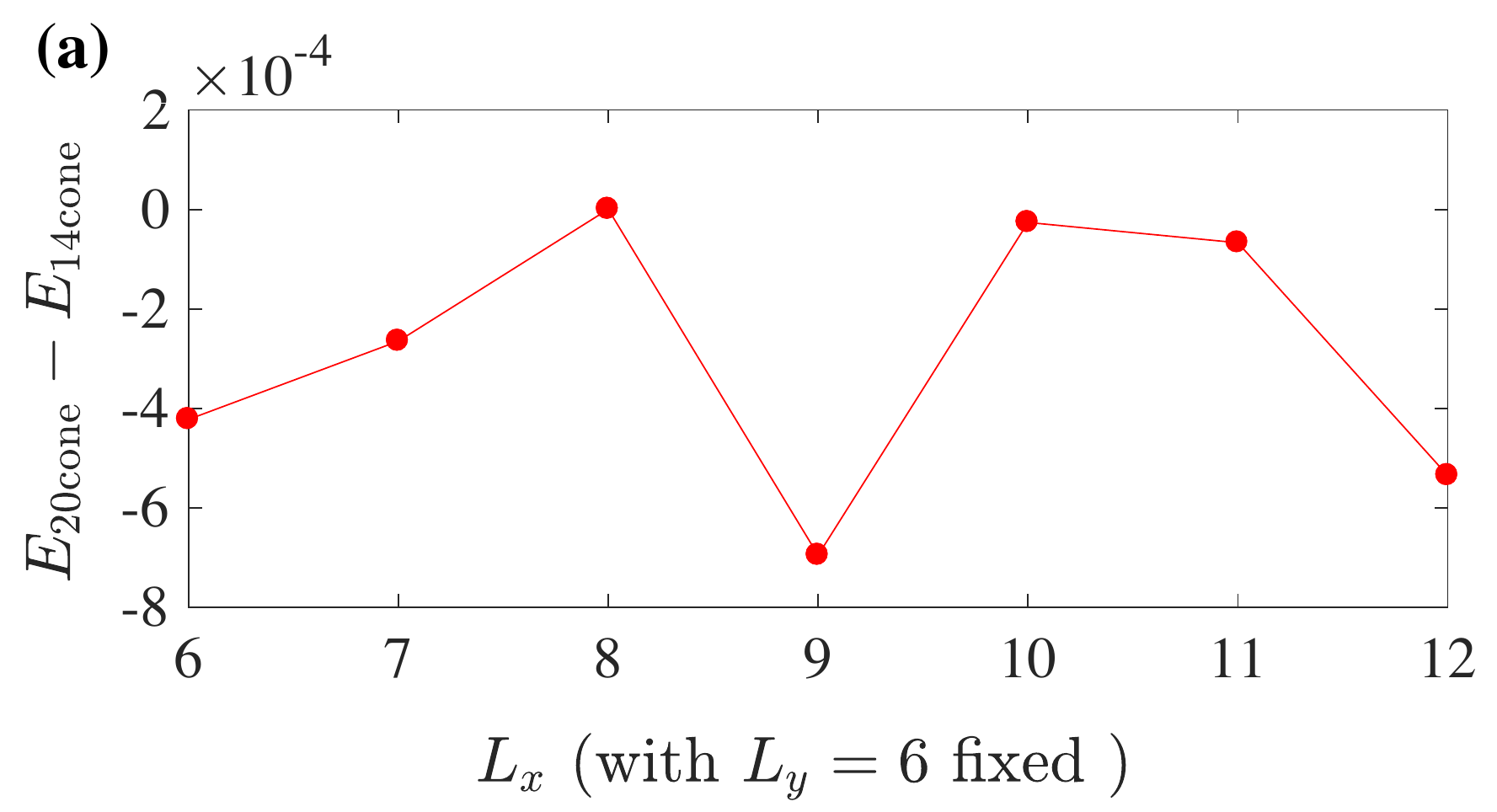}\\
\includegraphics[width=7. cm]{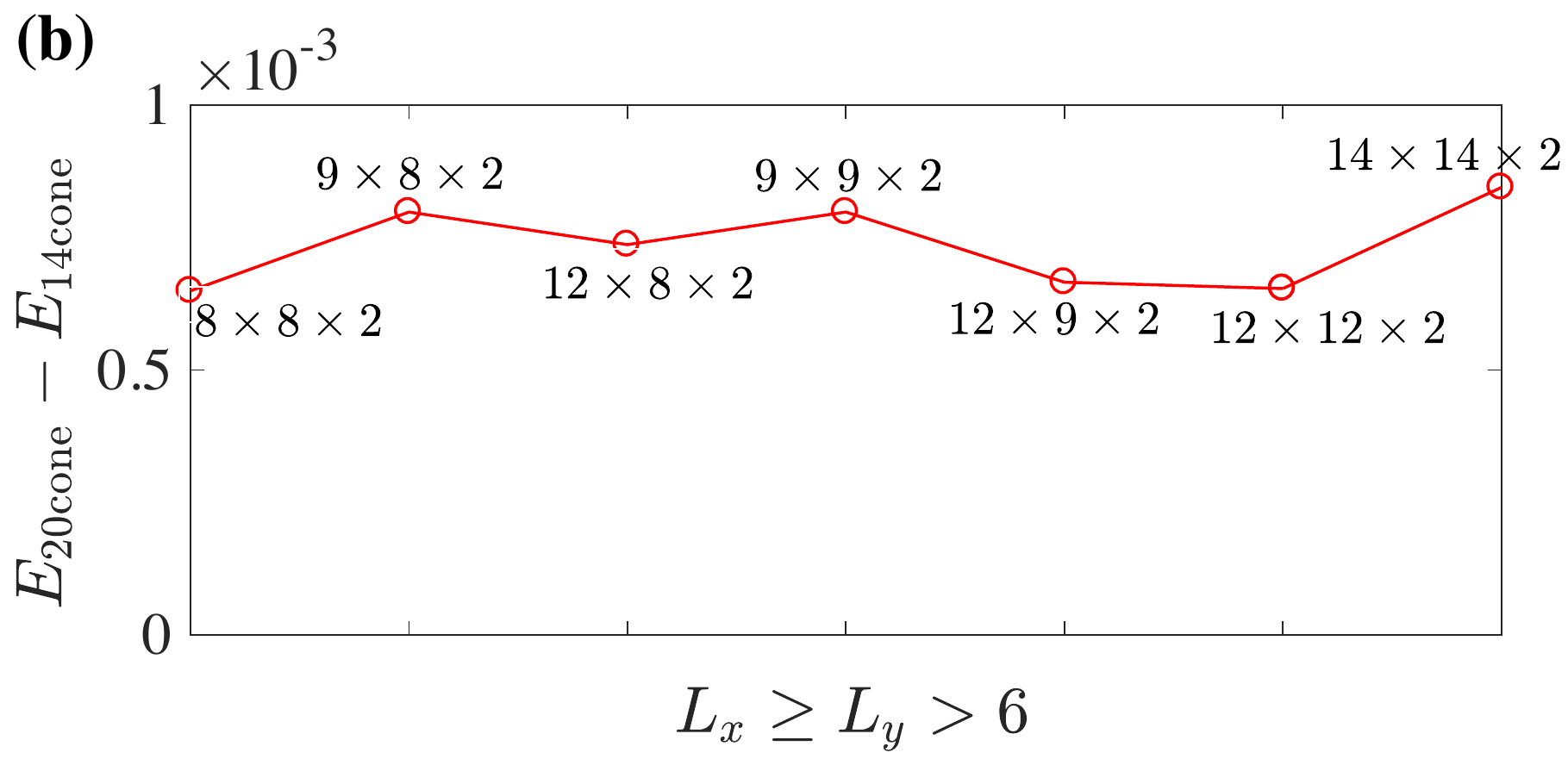}
\caption{(a) Energy difference per site between the 20-cone and 14-cone
states in VMC calculations for systems with short dimension $L_y = 6$, where
the 20-state lies lower in energy. (b) Energy difference for system sizes
$L_x$$\times$$L_y$$\times$2 with $L_x \geq L_y > 6$, where the 14-cone
state is lower in energy.}
\label{20VS14}
\end{figure}

Further investigation reveals, however, that the $L_y = 6$ geometry is still
anomalous, especially in the PKSL regime. In our VMC calculations for the
12$\times$6$\times$2 system we find that the ground state in the intermediate
regime $0.23 < \Gamma/|K| < 0.55$ is a gapless Z$_2$ QSL corresponding in the
2D system to the 20-cone state. This competes with the 14-cone
state, whose energy is higher by values ranging from 0 to 7$\times$$10^{-4}|K|$,
as shown in Fig.~\ref{20VS14}(a). However, for all larger accessible system
sizes with $L_x \geq L_y > 6$, these conditions are reversed and the 14-cone
state becomes lower in energy, as detailed in Fig.~\ref{20VS14}(b). Although
our system sizes preclude a reliable extrapolation to the thermodynamic limit,
by calculations on systems up to 14$\times$14$\times$2 we infer that the ground
state in this limit is indeed the 14-cone PKSL phase, and that the 20-cone
ground state is an artifact of the $L_y = 6$ geometry. Here we comment that the
energies we calculate for the 12$\times$12$\times$2 and 14$\times$14$\times$2
geometries are not obtained variationally, but are computed using the
variational parameters of the same states on the 12$\times$8$\times$2 cluster.
From these smaller systems we find that changes to the variational parameters
are negligible when compared to changes due to the cluster geometry,
and hence we infer that the large-system energies are reliable.

\begin{figure}[t]
\includegraphics[width=8.5cm]{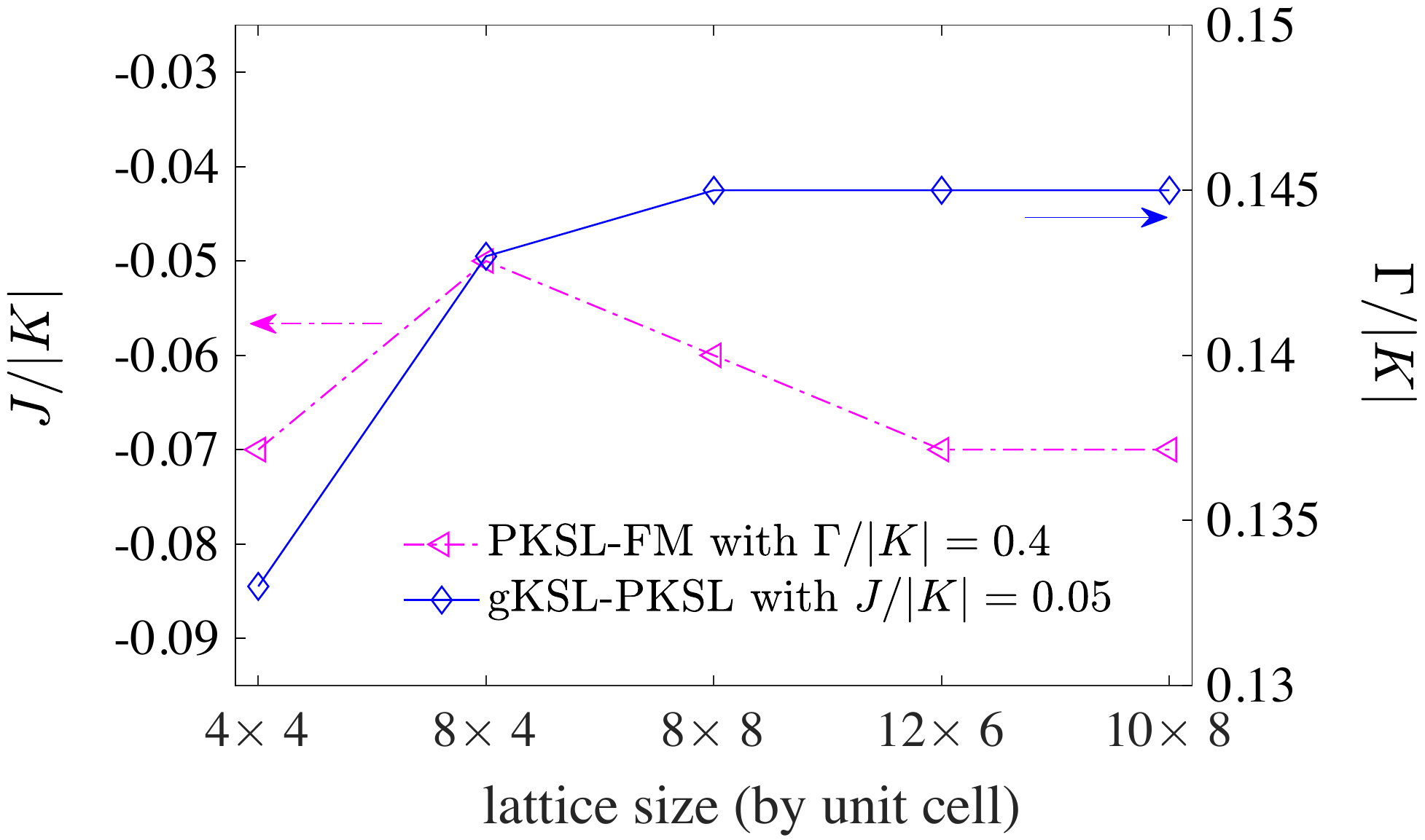}
\caption{Dependence on system size of the phase boundary between the gKSL and
PKSL phases, computed at fixed $J/|K| = 0.05$, and of the boundary between the
PKSL phase and the FM ordered phase, computed at fixed $\Gamma/|K| = 0.4$.}
\label{fsphsbdr}
\end{figure}

In view of all of these complexities, we performed a large number of studies
to ensure that the phase diagram presented in Fig.~1 of the main text is
completely reliable. To eliminate finite-size effects to the best of our
ability, we determined every point in the phase diagram by examining its
evolution with system size. We illustrate the results of these investigations
in Fig.~\ref{fsphsbdr}, where we show the size-dependence of the phase boundary
in two cases. For the transition from a QSL (the PKSL) to an ordered state
(the FM phase), we find rather little variation even from small sizes; for
the transition of most interest to the present study, that from the gKSL to
the PKSL phase, we find that a stable limiting value does require system
sizes large in both dimensions.

\section{Dynamical Structure Factor}\label{DSF}

\begin{figure}[t]
\includegraphics[width=6.cm]{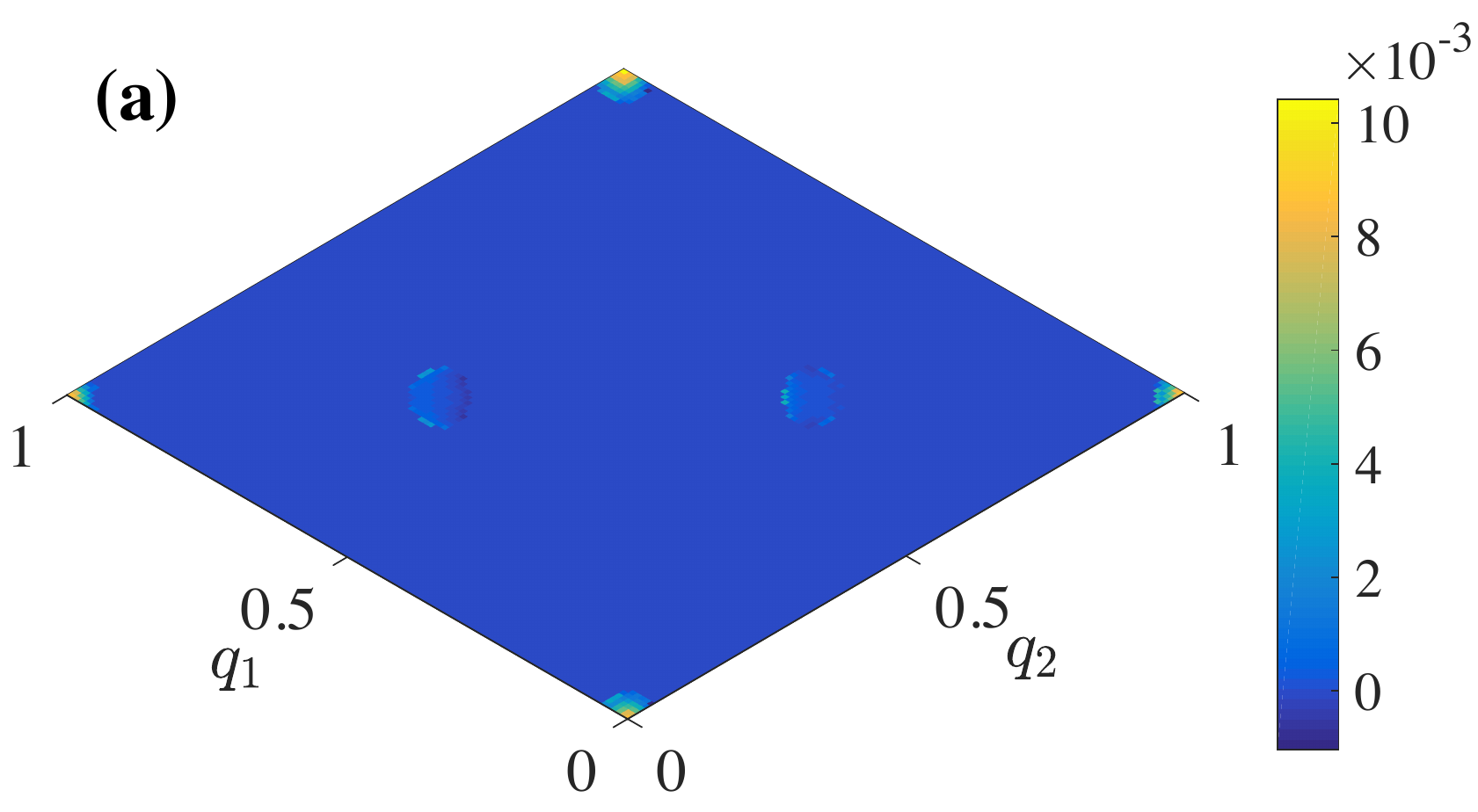}
\includegraphics[width=6.cm]{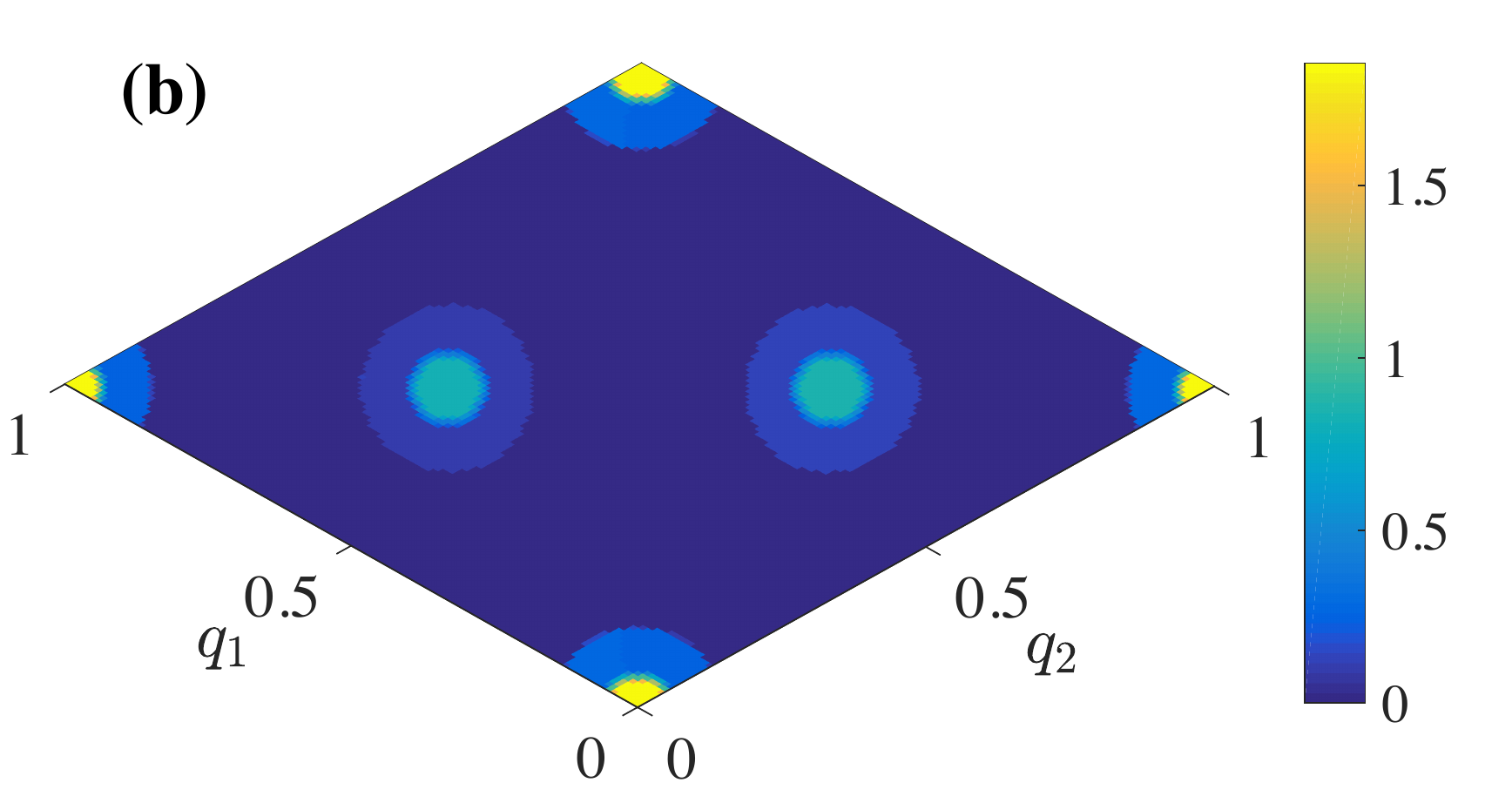}
\caption{Dynamical structure factor illustrated for the gKSL ($\Gamma/
|K| = 0.1$, $J/|K| = 0.05$). $\omega_0 = 0.082|K|$ is the minimum energy of
the gapped, $b^m$-dominated spinon bands. (a) Sum over all frequencies $\omega
\leq \omega_0$. (b) Sum over frequencies $\omega_0 \leq \omega \leq \omega_0
 + 0.018|K|$. Note the logarithmic intensity scale. }
\label{DSF_KSL}
\end{figure}

To understand the nature of the PKSL, we deduce its physical properties
and contrast them with those of the gKSL, which is also a gapless QSL. To
this end we investigate the magnetic response in the form of the dynamical
structure factor (DSF), $S^{mn}(\pmb q,\omega)$ ($m,n = x,y,z$), which we
compute at the level of linear response. From the fluctuation-dissipation
theorem, the DSF is defined by
\begin{equation}
S^{mn}(\pmb q,\omega) = - \frac{1}{\pi} [1 + n_B(\omega)] {\chi''}^{mn}
(\pmb q,\omega),
\end{equation}
where $n_B(\omega)$ is the Bose-Einstein distribution and ${\chi''}^{mn} (\pmb
q,\omega) = \lim_{\delta \to 0} \, {\rm Im\ } \chi^{mn} (\pmb q,\omega + i\delta)$
is the imaginary part of the analytic continuation of the finite-temperature
susceptibility, $\chi^{mn}(\pmb q,i\omega)$. To interpret the low-energy spin
response of the PKSL, it is most convenient to express the DSF in the Majorana
representation of Eq.~(\ref{emfd}). As an example, we illustrate the processes
contributing to $\chi^{zz} (\pmb q,i\omega)$ in the form
\Beq
& & \chi^{zz} (\pmb q,i\omega) = \int_0^\beta \langle T_\tau S^z_{\pmb q} (\tau)
S^z_{-\pmb q} (0) \rangle e^{i\omega \tau} d\tau \\
& & \ \ \ = - \int_0^\beta \sum_{\pmb k \pmb p} \langle T_\tau c_{-\pmb k} (\tau)
b^z_{\pmb k+\pmb q} (\tau) c_{-\pmb p} (0) b^z_{\pmb p-\pmb q} (0) \rangle e^{i\omega \tau}
d\tau \\
& & \ \ \ = \int_0^\beta \sum_{\pmb k} \left[ \langle T_\tau c_{-\pmb k}(\tau)
c_{\pmb k}(0) \rangle \langle b^z_{\pmb k+\pmb q} (\tau) b^z_{-\pmb k - \pmb q} (0)
\rangle  \right. \\
& & \left. \ \ \ \ \ \ \ \ \ \ \ \ \ -\langle T_\tau c_{-\pmb k} (\tau) b^z_{\pmb k}
(0) \rangle \langle b^z_{\pmb k+\pmb q} (\tau) c_{-\pmb k-\pmb q} (0) \rangle \right]
e^{i\omega \tau} d\tau,
\Eeq
where $T_\tau$ denotes time-ordering and $\langle \dots \rangle$ the thermal
average (which at zero temperature is equivalent to averaging over the
mean-field ground state). It is clear that processes contributing to the DSF
involve the excitation of two Majorana fermions created by a $c$ and a $b^z$
operator, with total momentum ${\pmb q}$ and total energy $\omega$. At low
energies, the DSF will be nonzero if one or both of the following two
situations is/are satisfied: (a) both the $c$ and the $b^z$ fermions are
gapless; (b) the $c$ and $b^z$ fermions are hybridized at the nodal points
of the PKSL dispersion.

\begin{table}[t]
\centering
\begin{tabular}{c|c|c||c|c|c|c}
\hline
\hline
$\frac{\Gamma}{|K|}$ & $\frac{J}{|K|}$ & Cone & $|\langle c_{\pmb k}|
\gamma_{\pmb k} \rangle|^2$ & $|\langle b^x_{\pmb k}|\gamma_{\pmb k} \rangle|^2$
& $|\langle b^y_{\pmb k}| \gamma_{\pmb k} \rangle|^2$ & $|\langle b^z_{\pmb k}|
\gamma_{\pmb k}\rangle|^2$ \\
\hline
 0    & 0.1  & K  & 1  & 0 & 0 & 0  \\
 \hline
 0.1  & 0.0  & K  & 1  & 0 & 0 & 0  \\
      & 0.05 & K  & 0.9997 & 1.04$\times$$10^{-4}$ & 1.04$\times$$10^{-4}$
      & 1.04$\times$$10^{-4}$  \\
\hline
     &     & K     & 0.3616 & 0.2128 & 0.2128 & 0.2128  \\
 0.3 & 0.0 & K$_1$ & 0.1850 & 0.4009 & 0.4009 & 0.0133 \\
     &     & K$_2$ & 0.2528 & 0.0717 & 0.0717 & 0.6038  \\
\hline
\hline
\end{tabular}
\caption{Projected weights of Majorana fermions at the gapless momentum
points in selected gKSL and PKSL phases when expressed in terms of the $c$,
$b^x$, $b^y$, and $b^z$ fermions [Eq.~(\ref{emfd})]. K$_1$ and K$_2$ denote
the first two blue points from left to right in Fig.~2(d) of the main text.}
\label{tab:Weights}
\end{table}

In the KSL ($J = 0 = \Gamma$), the $c$ fermions are gapless and independence
of the four species is a defining property, hence neither (a) nor (b) is
applicable and the spin response is gapped. In the gKSL, the $b^m$-dominated
bands retain a robust gap, which we label $\omega_0$ and which precludes case
(a). In Ref.~\cite{rsyb} it was argued qualitatively, on the basis of symmetry
and gauge constraints, that the system remains gapped when $J = 0$ or $\Gamma
 = 0$, but that the generic spin response is gapless. It is for this reason
that we distinguish between the KSL and the gKSL. In our spinon representation,
the nature of this case (b) gapless response is quantified through the
$c$-$b^m$ hybridization, which we compute at $\omega \rightarrow 0$ by
projecting the eigenstates we obtain onto the $\{c,b^x,b^y,b^z\}$ basis. The
results are shown in Table \ref{tab:Weights}: for the parameter sets
($\Gamma/|K| = 0.1$, $J = 0$) and ($\Gamma = 0$, $J/|K| = 0.1$), the ground
states we find are those with $\eta_3 = 0$ and $\eta_5 = 0$, ensuring that
hybridization is completely absent; by contrast, for ($\Gamma/|K| = 0.1$,
$J/|K| = 0.05$) the optimized values of $\eta_3$ and $\eta_5$ are nonzero but
remain small [O($10^{-2}$)], as a result of which the hybridization is very
weak [O($10^{-4}$)]. The corresponding DSF, shown in Figs.~\ref{DSF_KSL}(a)
and \ref{DSF_KSL}(b), remains strongly suppressed until finite energies
$\omega \approx \omega_0$. Thus although the gKSL has a gapless spin
response when both $J$ and $\Gamma$ are finite, in practice the degree of
hybridization remains small out to the first-order transitions terminating
its phase space, and hence the DSF may be very small at frequencies $\omega
 < \omega_0$. This situation could result in the erroneous identification of
a gapped QSL in inelastic neutron scattering experiments, which cannot access
the lowest energy scales, and may require nuclear magnetic or electron spin
resonance (NMR or ESR) measurements to identify the gapless nature. We
comment that the mean-field band gap of the $b^m$ fermions we compute is
larger than the true flux-excitation gap, as pointed out by Kitaev \cite{c0},
and thus the mean-field DSF provides only a qualitative reflection of the
spin response, but not a quantitative statement of the value $\omega_0$ that
governs it.

In the PKSL, the projected weights shown in Table \ref{tab:Weights}
demonstrate clearly that all of the eigenstates are strongly hybridized
superpositions of all four basis states, at all of the cones and even as
$\omega \rightarrow 0$. Thus both (a) and (b) are relevant. The low-energy
DSF shown in Fig.~2(e) of the main text is a consequence of two-cone
processes involving any of the 14 cones, with different weights for each
different type of intra- or intercone pair. When the energy increases towards
the width of the lowest band, the DSF is no longer controlled by the nodal
points and becomes continuous across the Brillouin zone, as shown in Fig.~2(f)
of the main text. At all energies the DSF is strongest at ${\pmb q} = 0$, and
indeed we note that the peaks in Fig.~2(e) of the main text have been cut from
an intensity of 16 to 6 in order to show the structure of the low-energy DSF
more clearly throughout the Brillouin zone.
\begin{figure}[t]
\includegraphics[width=6.5cm]{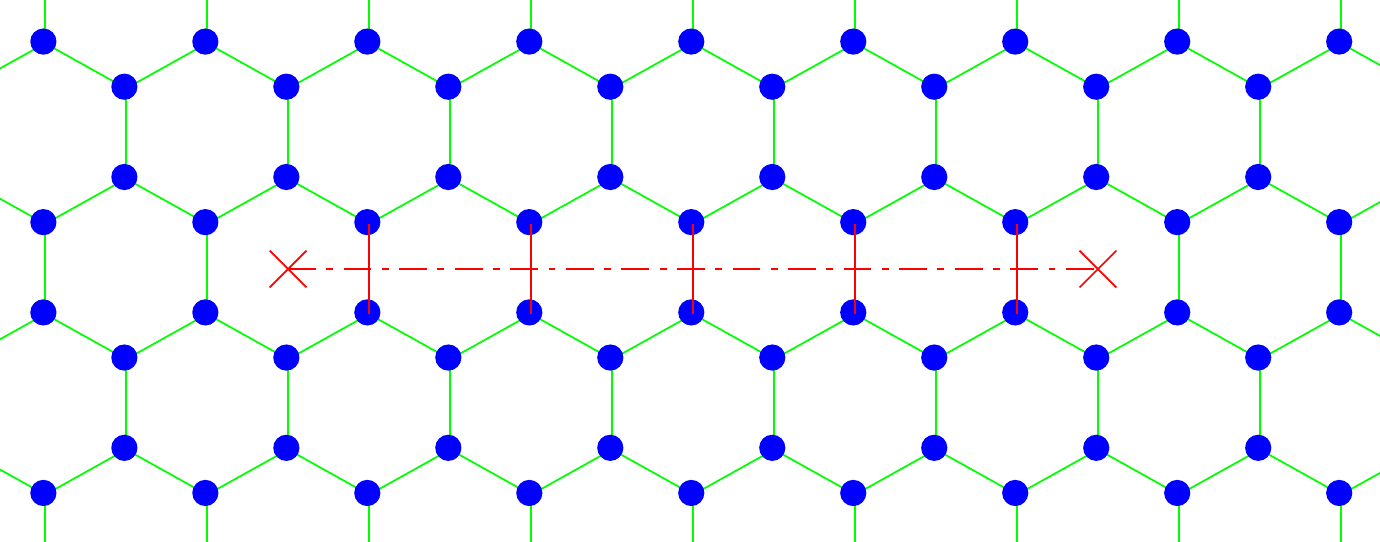}
\caption{Two vortices inserted at the hexagons marked by red crosses
define a branch cut (dot-dashed line).}
\label{Vortices}
\end{figure}

\section{Field-Induced Topological States}\label{SLinField}

As we have shown in the main text, the PKSL in an applied magnetic field
$\pmb B \parallel {\hat c}$ provides examples of the $\nu = 5$ and $\nu = 4$
chiral spin-liquid (CSL) states of Ref.~\cite{c0}. While high-$\nu$ CSL phases
have been studied in SO($n$)-symmetric \cite{rt,rywtsc} and spin $S > 1/2$
systems \cite {Thomale09,S=1NCSL}, here we have realized them in a $S = 1/2$
lattice model with only short-range interactions. In an in-plane field
$\pmb B \parallel (\pmb x - \pmb y)$, the gapless 14-cone state is preserved
for small but finite field strengths. The phase diagrams as a function of
$|\pmb B|$ are shown in Fig.~3 of the main text and here we provide the
results supporting our conclusions.

\begin{figure*}[t]
\includegraphics[width=4.4cm]{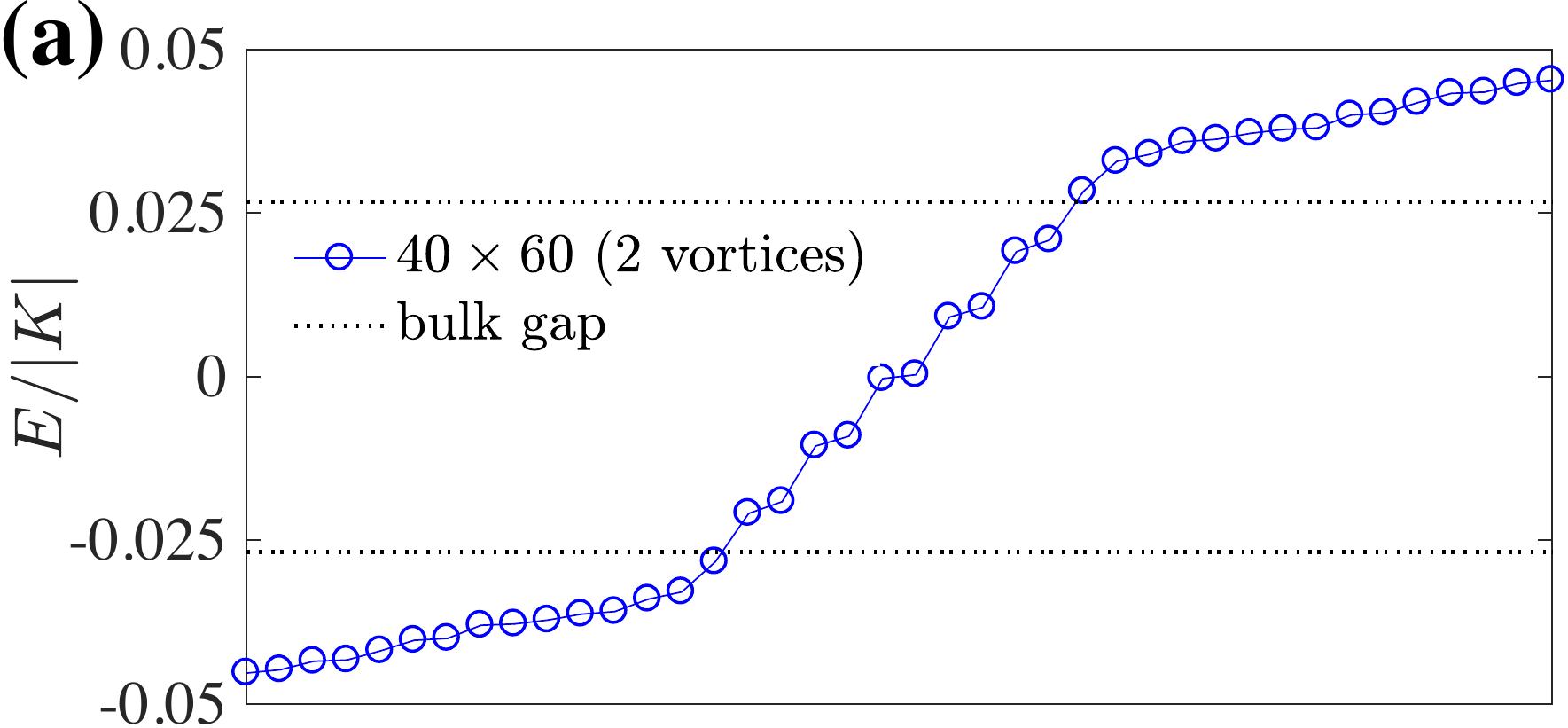}\
\includegraphics[width=4.4cm]{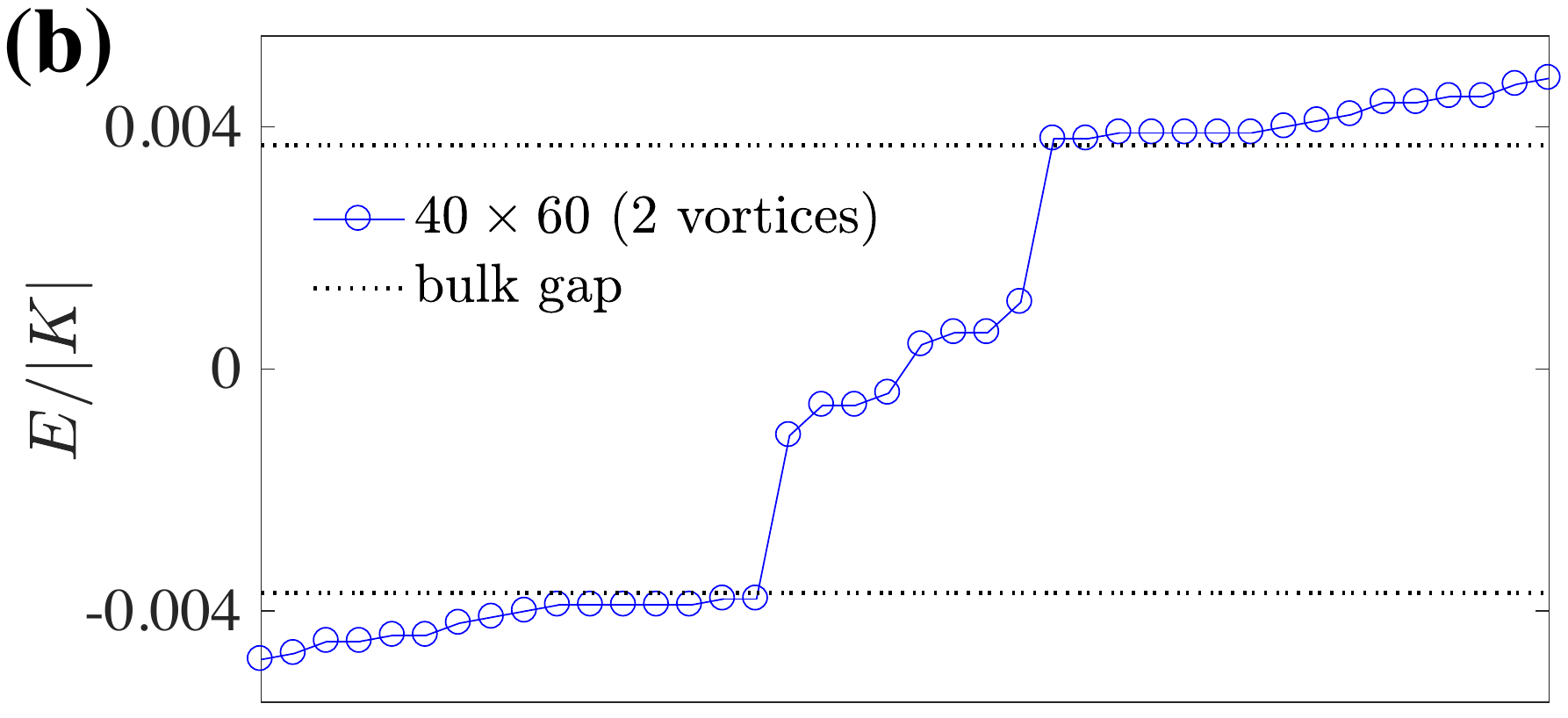}\
\includegraphics[width=4.4cm]{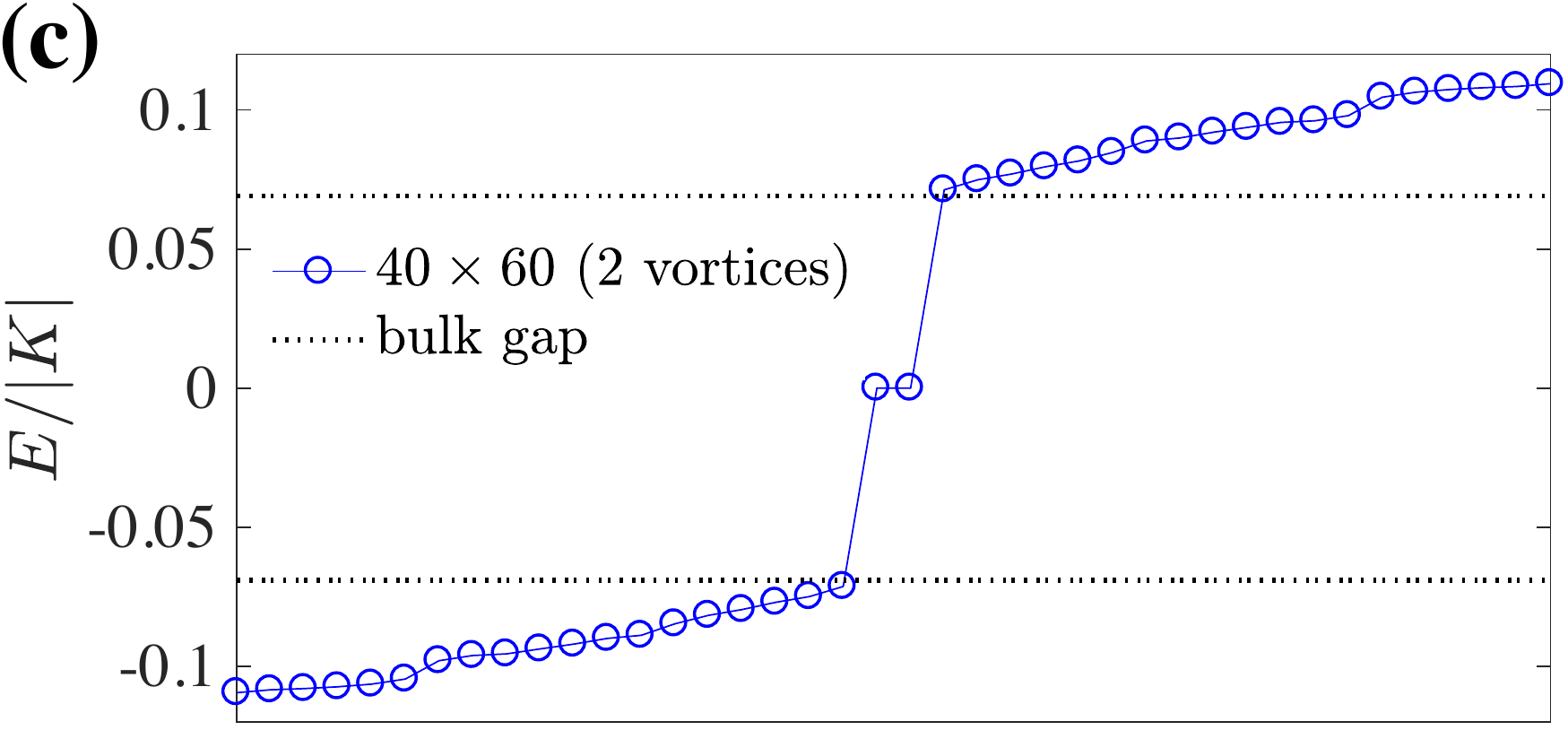}\
\includegraphics[width=4.4cm]{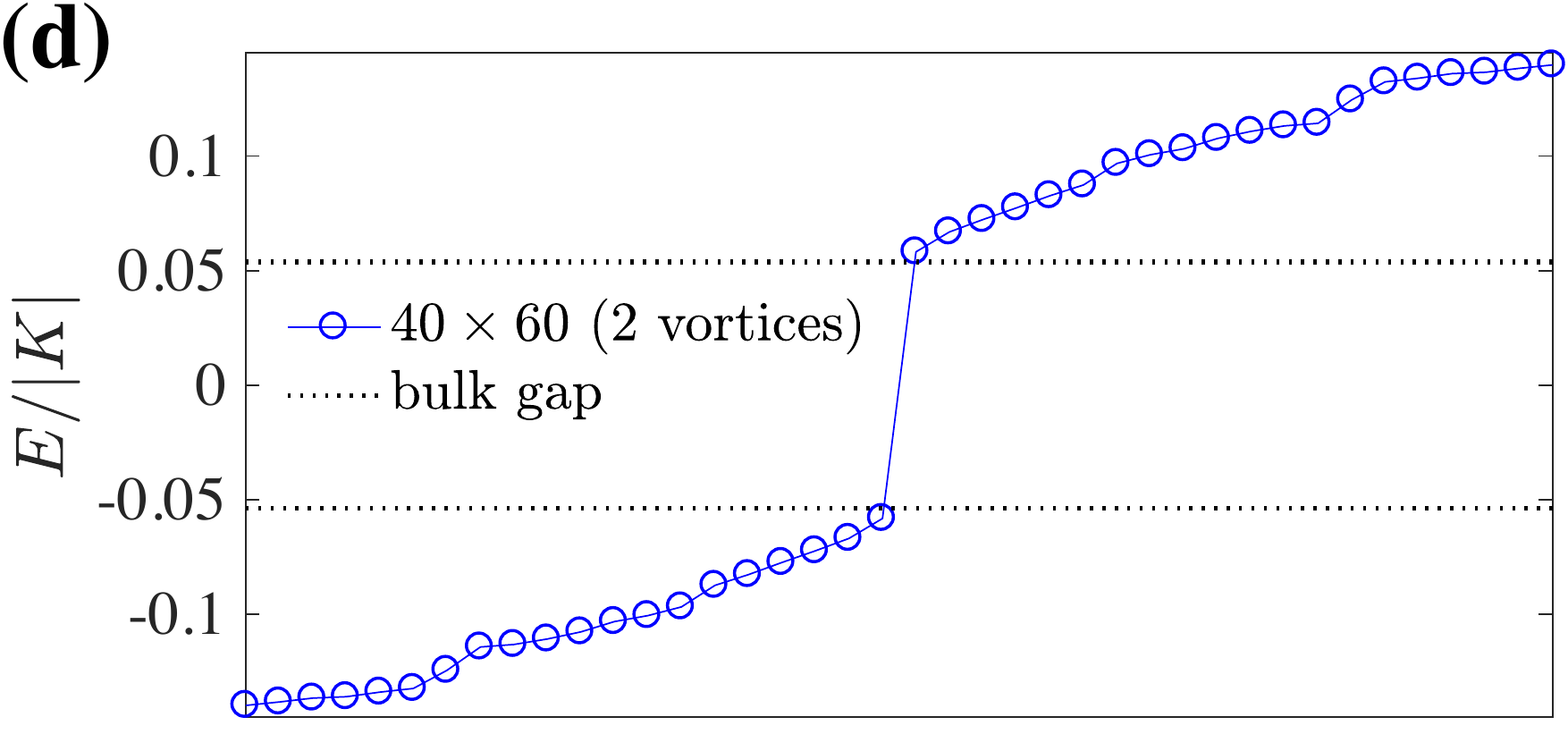}
\caption{Energy spectrum of the mean-field Hamiltonian in the vicinity of
the gap, computed for a large system in four different external magnetic
fields $\pmb B \parallel \pmb c$ and in the presence of two vortices
(Fig.~\ref{Vortices}). The interaction parameters are those of the PKSL
($\Gamma = 0.3|K|$, $J = 0$) and the mean-field parameters are determined
by VMC calculations with system size 8$\times$8$\times$2. (a) $\tilde B_x =
0.1$, where each vortex binds two complex fermion modes plus a Majorana zero
mode ($\nu = 5$); (b) $\tilde B_x = 0.4$, where each vortex binds two mid-gap
complex fermion modes ($\nu = 4$); (c) $\tilde B_x = 0.5$, where each vortex
binds a Majorana zero mode ($\nu = 1$; this state becomes trivial after
Gutzwiller projection); (d) $\tilde B_x = 0.85$, where no mid-gap modes
are trapped by the vortex.}
\label{ZMs}
\end{figure*}

To perform VMC calculations in a magnetic field, we add the Zeeman-coupling
term,
\Beq
H_{\rm Zeeman} = g \mu_B \pmb B \cdot {\textstyle \frac12} C^\dag_i {\pmb {\sigma}}
C_i,
\Eeq
to the mean-field Hamiltonian and reoptimize the parameters under Gutzwiller
projection. For convenience we introduce the normalized field strength
$\tilde B = g \mu_B B/|K|$. As in Ref.~\cite{c2}, we assume an uniform and
isotropic $g$-factor for all field directions. To investigate the physics of
anyons in the CSL phases, we introduce a pair of vortices in the system, as
represented in Fig.~\ref{Vortices}, by reversing the signs of the mean-field
terms on the bonds intersected by the dot-dashed line.

\subsection{$\pmb B \parallel \pmb c$}

In the gKSL for $K < 0$, there is a single phase transition with
increasing the field strength \cite{HCJ}. For $\tilde B < {\tilde B}_c$
($= 0.19$ for ${\Gamma/|K|} = 0.1$, $J = 0$), the mean-field dispersion is
fully gapped, this gap growing linearly with $|\tilde B|$ when $|\tilde B|$
is small, except when $J = 0$ or $\Gamma = 0$, in which case it is proportional
to $|\tilde B|^3$. The ground state has Chern number $\nu = 1$, so one Majorana
zero mode is bound to each vortex, $\sigma$, and the vortices obey the
well-known fusion rule $\sigma \times \sigma = 1 + \varepsilon$, where $1$
is the vacuum and $\varepsilon$ is a fermion. $1$, $\varepsilon$, and
$\sigma$ are the three topologically distinct sectors in the system, as
shown in Table I of the main text, and correspond to the GSD of 3.

To show this we apply the following considerations. A fermion experiences a
flux $\pi$ on moving around a vortex, whence the vortices appear in pairs. A
pair of $\pi$-fluxes gives rise to two Majorana zero modes, equivalent
to a single fermion mode, whose energies fall in the bulk gap [similar to
Fig.~\ref{ZMs}(c)]. Because of the nonzero Chern number, each vortex
($\sigma$) is associated with a gauge charge 1/4, a situation analogous to
the Hall effect \cite{Nayak_RMP}. When two vortices are exchanged, the
Aharonov-Bohm phase between the $\pi$-flux and the associated charge is counted
only once. Thus the topological spin (the Abelian phase obtained by rotating a
vortex by $2\pi$, which by the spin-statistics theorem is the same as the phase
obtained by exchanging two vortices) is equal to $e^{i{\pi\over4} \times {1\over2}} =
e^{i{\pi\over8}}$.

If a fermion, which has gauge charge $-1$ \cite{c0}, is attached to a vortex,
the composite vortex $\sigma \times \varepsilon$ also has topological spin
$e^{i{\pi\over8}}$: exchanging two composite vortices creates a phase
$e^{i({\pi\over8}-{\pi\over2}-{\pi\over2}+\pi)}$, where the factors of $e^{-i{\pi\over2}}$
arise from the fermion and $\pi$-flux moving half a circle around each other
and the phase $e^{i\pi}$ from the exchange of two fermions. In fact a composite
vortex has the same properties as a vortex (except for the gauge charge, which
is not measurable), resulting in the fusion rule $\sigma \times \varepsilon =
\sigma$ \cite{c0}, confirming that the set of three topologically distinct
sectors is closed.

After Gutzwiller projection, if it remains Z$_2$-deconfined then the $\nu = 1$
mean-field state becomes a non-Abelian CSL, in which the vortices are Ising
anyons. Our VMC calculations for the gKSL in a field verify that the GSD of
this CSL on a torus is 3, as stated in Sec.~\ref{IGGPSG} for any system with
an odd Chern number. When $\tilde B > {\tilde B}_c$, the projected state
is Z$_2$-confined and the GSD is 1, i.e.~the system enters a topologically
trivial gapped phase, which is connected adiabatically to the fully polarized
state.

Turning to the PKSL, one may apply all of the same considerations. We find
that there are two phase transitions, which for $\Gamma/|K| = 0.3$, occur at
$\tilde B_1 = 0.28$ and $\tilde B_2 = 0.71$. When $\tilde B < \tilde B_1$, the
Chern number is $\nu = 5$: the vortex carries gauge charge $\nu/4 =
5/4$, which give rise to a topological spin $e^{i{5\over8}\pi}$. One vortex
traps two complex fermion modes plus one unpaired Majorana zero mode, as shown
in Fig.~\ref{ZMs}(a) by computing the low-energy spectrum at the mean-field
level for a large system. The odd Chern number means that the vortex is a
non-Abelian anyon and correspondingly the $\nu = 5$ phase is a non-Abelian CSL.
This nontrivial topological order is verified in VMC by the three-fold
degeneracy of the ground state on a torus, as shown by the eigenvalues of
the density matrix listed in Table \ref{tab:GSD}. Despite their commonalities,
the $\nu = 1$ and $\nu = 5$ phases differ in the topological spin of their
vortices, their total chiral central charge, and their thermal Hall
conductance (Table I of the main text).

\begin{figure*}[t]
\includegraphics[width=5.9cm]{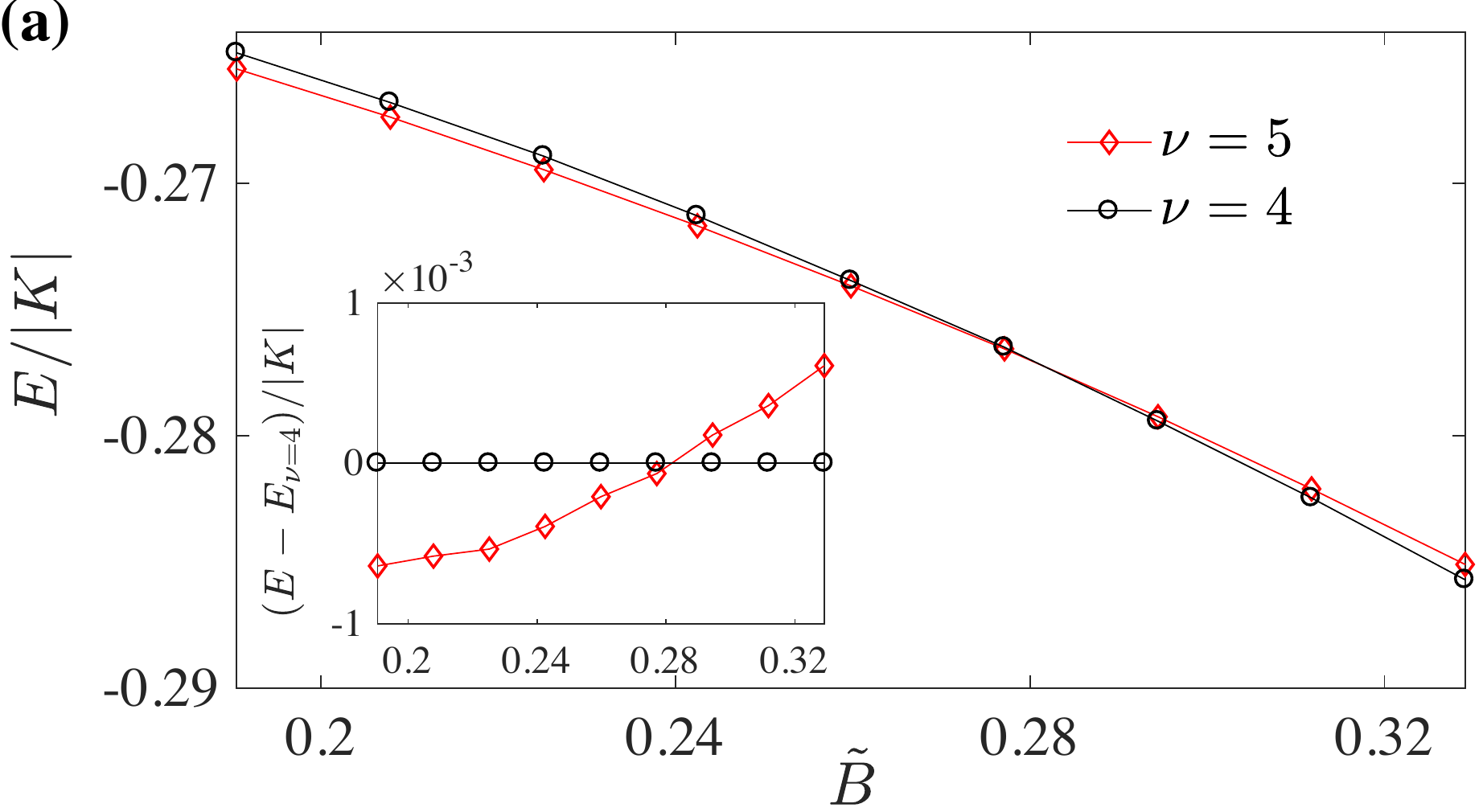}
\includegraphics[width=5.9cm]{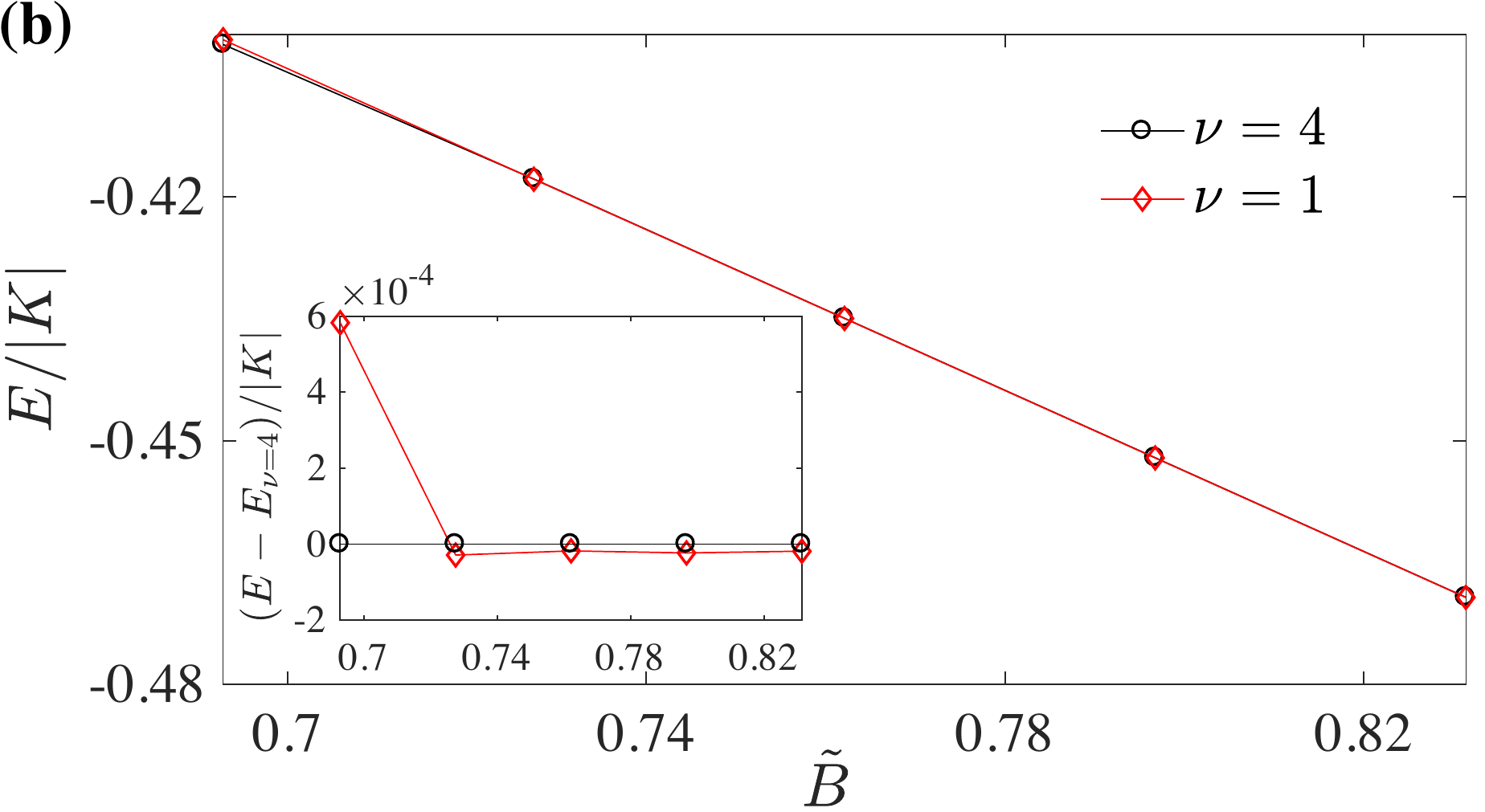}
\includegraphics[width=5.9cm]{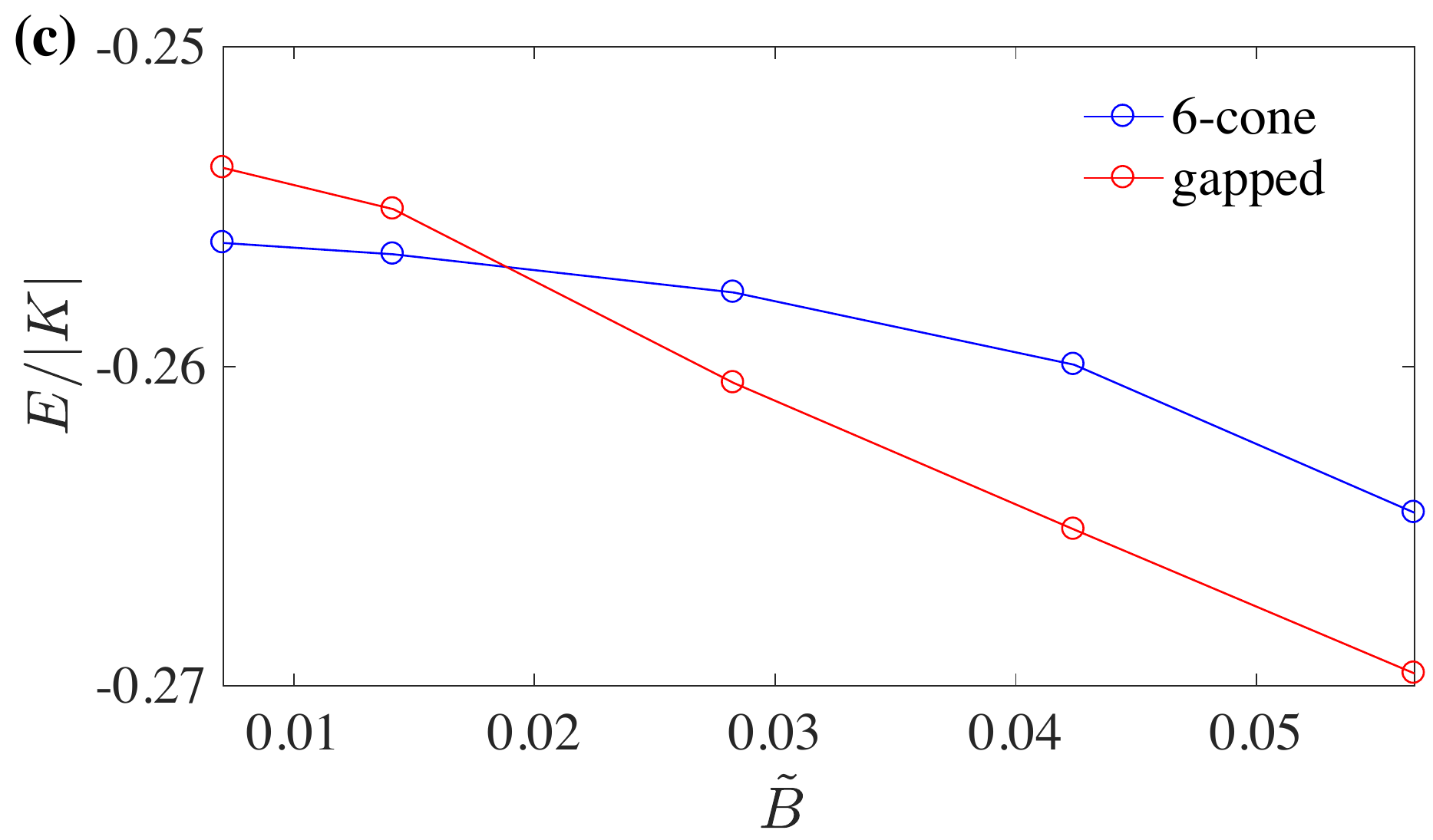}
\caption{Field-induced phase transitions in the PKSL ($\Gamma/|K| = 0.3$,
$J = 0$). (a) Level-crossing at the transition from the $\nu = 5$ to the
$\nu = 4$ CSL in a field $\pmb B \parallel \pmb c$ and (b) level-crossing
at the transition from the $\nu = 4$ CSL, which is Z$_2$-deconfined, to
the trivial (Z$_2$-confined) phase obtained by projection of the $\nu = 1$
mean-field state. Both insets show the energy difference per site with
respect to the $\nu = 4$ state. (c) Level-crossing from the gapless 6-cone
state to a trivial gapped phase in a field $\pmb B \parallel (\pmb x -
\pmb y)$.}
\label{levelcross}
\end{figure*}

\begin{table}[b]
\centering
\begin{tabular}{c|c||c|c|c|c|c}
\hline
\hline
$\tilde B_x$ & $\, \nu \,$ & $\rho_1$ & $\rho_2$ & $\rho_3$ & ($\rho_4$) & GSD \\
\hline
0.1  & $5$  & 0.4016   & 0.8456  & 1.7527   &        & 3  \\
\hline
0.4  & $4$  & 0.7506   & 0.8168  & 0.8303   & 1.6024 & 4 \\
\hline
0.5 &$ 1$  & $\,$8.0$\times$10$^{-6}$ & $\,$7.6$\times$10$^{-4}$ & 2.9992  &
& 1 \\
\hline
0.85 & $0$  & $\,$7.5$\times$10$^{-9}$ & $\,$3.0$\times$10$^{-7}$ &
$\,$5.6$\times$10$^{-6}$ & $\,$4.0000$\,$ & 1 \\
\hline
\hline
\end{tabular}
\caption{Eigenvalues of the density matrices of the CSL ground states induced
by a magnetic field $\pmb B \parallel \hat c$, computed in the PKSL ($\Gamma/
|K| = 0.3$, $J = 0$) for a system of size 8$\times$8$\times$2. $\nu$ is the
mean-field Chern number; as discussed in the text, the projected $\nu = 1$
and $\nu = 0$ states are the same phase.}
\label{tab:GSD}
\end{table}

When $\tilde B_1 < \tilde B < \tilde B_2$, the Chern number is $\nu = 4$. As
shown in  Fig.~\ref{ZMs}(b), a pair of vortices gives rise to four mid-gap
fermion modes, whose energies are nonzero. In principle each vortex 
(denote by $m$ \cite{c0}) traps four Majorana zero modes, but these lower their
energies by coupling to form two complex fermion modes. Thus the vortex, $m$,
carries gauge charge $\nu/4 = 1$ and has topological spin $e^{i{\pi\over2}\times{1}}
 = e^{i{\pi\over2}}$, meaning that it is a semion. The composite object of two
$m$ vortices has statistical angle $\pi/2 \times 4 = 2 \pi$, which means
it is a boson ($m \times m = 1$).

The attachment of a fermion, $\varepsilon$, to the vortex creates a new
composite vortex, $m \times \varepsilon = e$. Similar to the previous
discussion, this does not change the topological spin, i.e.~$e$ is also a
semion, with $e \times e = 1$. Thus it is easy to obtain the remaining
fusion rules, $e \times m = \varepsilon$, $e \times \varepsilon
 = m$, and $m \times \varepsilon = e$. The nontrivial topological order
of the projected $\nu = 4$ state is reflected by the GSD on a torus, which is
4 (Table \ref{tab:GSD}).

For $\tilde B > \tilde B_2$, the outcome of VMC is a trivial, gapped,
Z$_2$-confined phase. However, working at the mean-field level (with the
parameters determined from VMC), there is an additional critical point at
$\tilde B_3 \sim 1.22$ where the mean-field Chern number changes from $\nu
 = 1$ to $\nu = 0$. The number of mid-gap states in the mean-field spectrum
changes correspondingly from 2 [Fig.~\ref{ZMs}(c)] to 0 [Fig.~\ref{ZMs}(d)].
However, as Table \ref{tab:GSD} shows clearly, both the projected $\nu = 1$
state and the projected $\nu = 0$ state have a GSD of 1, showing that they
are Z$_2$-confined. Usually, a mean-field state with nontrivial Chern number
is nontrivial after Gutzwiller projection, but in the present case our
numerical results indicate that the strong magnetic field causes the highly
polarized projected state to become Z$_2$-confined. This confinement can be
interpreted as a closing of the Z$_2$ vison gap despite the fact that the
mean-field spinon gap is still finite. We comment in closing this subsection
that the transitions at both $\tilde B_1$ and $\tilde B_2$ are first-order,
as shown in Figs.~\ref{levelcross}(a) and \ref{levelcross}(b), and hence the
closure of the spinon gap required at a continuous topological transition is
absent.

\subsection{$\pmb B \parallel (\pmb x - \pmb y)$}

Here we discuss only the PKSL. An in-plane magnetic field breaks the
three-fold rotation symmetry and in general a gap will be opened at all the
cones. However, as noted in the main text, for the special directions $\pmb B
\parallel (\pmb x - \pmb y)$, $(\pmb y - \pmb z)$, and $(\pmb z-\pmb x)$, 6
of the cones remain gapless at small but finite fields. As the field is
increased, pairs of cones move together and may merge, then become gapped at
a second-order phase transition \cite{c2}. In the PKSL, a first-order
transition preempts this process, as shown in Fig.~\ref{levelcross}(c), and
instead the system enters a phase whose GSD on a torus is 1, meaning that it
is Z$_2$-confined and trivial. We find that this state, which is connected
directly to the fully polarized limit, is characterized by strong spatial
anisotropy and a sign-reversal of the variational parameter $\rho_a$.
All of the first-order transitions in Fig.~\ref{levelcross} reinforce the
observation that many states compete strongly at low energies, which in turn
raises the computational complexity of the task of deducing the full phase
diagram in $K$, $J$, $\Gamma$, $|\pmb B|$, and $\hat B$.

\subsection{General $\pmb B$}

Before concluding this section, we provide one brief illustration of the
possibilities that arise when applying the magnetic field in an intermediate
direction, and when applying it to an ordered phase. For a field orientation
fixed along the direction $[1,1,0.2]$ and interaction parameters fixed at
(${\Gamma/|K|} = 1.4$, $J = 0$) \cite{c2}, we observe a phase transition at
$\tilde B = 0.36$ from the zigzag phase to a $\nu = 1$ CSL. This CSL phase
is stable in the field range $0.36 < \tilde B < 0.50$. It supports a
half-quantized thermal Hall conductance, a result which may be of key
importance in interpreting the results of experiments such as that reported
in Ref.~\cite{ThermalHall}.